
%
%
%
\def\unredoffs{} \def\redoffs{\voffset=-.31truein\hoffset=-.59truein}
\def\speclscape{\special{ps: landscape}}
%
%
%
%
\newbox\leftpage \newdimen\fullhsize \newdimen\hstitle \newdimen\hsbody
\tolerance=1000\hfuzz=2pt
\catcode`\@=11 
\def\bigans{b }
\def\answ{b }
\ifx\answ\bigans\message{(This will come out unreduced.}
\magnification=1200\unredoffs\baselineskip=16pt plus 2pt minus 1pt
\hsbody=\hsize \hstitle=\hsize 
\else\message{(This will be reduced.} \let\l@r=L
\magnification=1000\baselineskip=16pt plus 2pt minus 1pt \vsize=7truein
\redoffs \hstitle=8truein\hsbody=4.75truein\fullhsize=10truein\hsize=\hsbody
\output={\ifnum\pageno=0 
  \shipout\vbox{\speclscape{\hsize\fullhsize\makeheadline}
    \hbox to \fullhsize{\hfill\pagebody\hfill}}\advancepageno
  \else
  \almostshipout{\leftline{\vbox{\pagebody\makefootline}}}\advancepageno
  \fi}
\def\almostshipout#1{\if L\l@r \count1=1 \message{[\the\count0.\the\count1]}
      \global\setbox\leftpage=#1 \global\let\l@r=R
 \else \count1=2
  \shipout\vbox{\speclscape{\hsize\fullhsize\makeheadline}
      \hbox to\fullhsize{\box\leftpage\hfil#1}}  \global\let\l@r=L\fi}
\fi
%
\newcount\yearltd\yearltd=\year\advance\yearltd by -1900

\def\Title#1#2{\nopagenumbers\abstractfont\hsize=\hstitle\rightline{#1}%
\vskip 1in\centerline{\titlefont #2}\abstractfont\vskip .5in\pageno=0}
\def\Date#1{\vfill\leftline{#1}\tenpoint\supereject\global\hsize=\hsbody%
\footline={\hss\tenrm\folio\hss}}
%

\def\draftmode{\message{ DRAFTMODE }\def\draftdate{{\rm preliminary draft:
\number\month/\number\day/\number\yearltd\ \ \hourmin}}%
\headline={\hfil\draftdate}\writelabels\baselineskip=20pt plus 2pt minus 2pt
 {\count255=\time\divide\count255 by 60 \xdef\hourmin{\number\count255}
  \multiply\count255 by-60\advance\count255 by\time
  \xdef\hourmin{\hourmin:\ifnum\count255<10 0\fi\the\count255}}}
\def\nolabels{\def\wrlabeL##1{}\def\eqlabeL##1{}\def\reflabeL##1{}}
\def\writelabels{\def\wrlabeL##1{\leavevmode\vadjust{\rlap{\smash%
{\line{{\escapechar=` \hfill\rlap{\sevenrm\hskip.03in\string##1}}}}}}}%
\def\eqlabeL##1{{\escapechar-1\rlap{\sevenrm\hskip.05in\string##1}}}%
\def\reflabeL##1{\noexpand\llap{\noexpand\sevenrm\string\string\string##1}}}
\nolabels
%
\global\newcount\secno \global\secno=0
\global\newcount\meqno \global\meqno=1
\def\newsec#1{\global\advance\secno by1\message{(\the\secno. #1)}
\global\subsecno=0\eqnres@t\noindent{\bf\the\secno. #1}
\writetoca{{\secsym} {#1}}\par\nobreak\medskip\nobreak}
\def\eqnres@t{\xdef\secsym{\the\secno.}\global\meqno=1\bigbreak\bigskip}
\def\sequentialequations{\def\eqnres@t{\bigbreak}}\xdef\secsym{}
\global\newcount\subsecno \global\subsecno=0
\def\subsec#1{\global\advance\subsecno by1\message{(\secsym\the\subsecno.
#1)}
\ifnum\lastpenalty>9000\else\bigbreak\fi
\noindent{\it\secsym\the\subsecno. #1}\writetoca{\string\quad
{\secsym\the\subsecno.} {#1}}\par\nobreak\medskip\nobreak}
\def\appendix#1#2{\global\meqno=1\global\subsecno=0\xdef\secsym{\hbox{#1.}}
\bigbreak\bigskip\noindent{\bf Appendix #1. #2}\message{(#1. #2)}
\writetoca{Appendix {#1.} {#2}}\par\nobreak\medskip\nobreak}
%
%
\def\eqnn#1{\xdef #1{(\secsym\the\meqno)}\writedef{#1\leftbracket#1}%
\global\advance\meqno by1\wrlabeL#1}
\def\eqna#1{\xdef #1##1{\hbox{$(\secsym\the\meqno##1)$}}
\writedef{#1\numbersign1\leftbracket#1{\numbersign1}}%
\global\advance\meqno by1\wrlabeL{#1$\{\}$}}
\def\eqn#1#2{\xdef #1{(\secsym\the\meqno)}\writedef{#1\leftbracket#1}%
\global\advance\meqno by1$$#2\eqno#1\eqlabeL#1$$}
%
\newskip\footskip\footskip14pt plus 1pt minus 1pt 
\def\footnotefont{\ninepoint}\def\f@t#1{\footnotefont #1\@foot}
\def\f@@t{\baselineskip\footskip\bgroup\footnotefont\aftergroup\@foot\let\next}
\setbox\strutbox=\hbox{\vrule height9.5pt depth4.5pt width0pt}
\global\newcount\ftno \global\ftno=0
\def\foot{\global\advance\ftno by1\footnote{$^{\the\ftno}$}}
%
\newwrite\ftfile
\def\footend{\def\foot{\global\advance\ftno by1\chardef\wfile=\ftfile
$^{\the\ftno}$\ifnum\ftno=1\immediate\openout\ftfile=foots.tmp\fi%
\immediate\write\ftfile{\noexpand\smallskip%
\noexpand\item{f\the\ftno:\ }\pctsign}\findarg}%
\def\footatend{\vfill\eject\immediate\closeout\ftfile{\parindent=20pt
\centerline{\bf Footnotes}\nobreak\bigskip\input foots.tmp }}}
\def\footatend{}
%
%
\global\newcount\refno \global\refno=1
\newwrite\rfile
\def\ref{[\the\refno]\nref}
\def\nref#1{\xdef#1{[\the\refno]}\writedef{#1\leftbracket#1}%
\ifnum\refno=1\immediate\openout\rfile=refs.tmp\fi
\global\advance\refno by1\chardef\wfile=\rfile\immediate
\write\rfile{\noexpand\item{#1\ }\reflabeL{#1\hskip.31in}\pctsign}\findarg}
\def\findarg#1#{\begingroup\obeylines\newlinechar=`\^^M\pass@rg}
{\obeylines\gdef\pass@rg#1{\writ@line\relax #1^^M\hbox{}^^M}%
\gdef\writ@line#1^^M{\expandafter\toks0\expandafter{\striprel@x #1}%
\edef\next{\the\toks0}\ifx\next\em@rk\let\next=\endgroup\else\ifx\next\empty%
\else\immediate\write\wfile{\the\toks0}\fi\let\next=\writ@line\fi\next\relax}}
\def\striprel@x#1{} \def\em@rk{\hbox{}}
\def\lref{\begingroup\obeylines\lr@f}
\def\lr@f#1#2{\gdef#1{\ref#1{#2}}\endgroup\unskip}
\def\semi{;\hfil\break}
\def\addref#1{\immediate\write\rfile{\noexpand\item{}#1}} 
\def\startrefs#1{\immediate\openout\rfile=refs.tmp\refno=#1}
\def\xref{\expandafter\xr@f}\def\xr@f[#1]{#1}
\def\refs#1{\count255=1[\r@fs #1{\hbox{}}]}
\def\r@fs#1{\ifx\und@fined#1\message{reflabel \string#1 is undefined.}%
\nref#1{need to supply reference \string#1.}\fi%
\vphantom{\hphantom{#1}}\edef\next{#1}\ifx\next\em@rk\def\next{}%
\else\ifx\next#1\ifodd\count255\relax\xref#1\count255=0\fi%
\else#1\count255=1\fi\let\next=\r@fs\fi\next}
%

%
\newwrite\ffile\global\newcount\figno \global\figno=1
\def\fig{fig.~\the\figno\nfig}
\def\nfig#1{\xdef#1{Fig.~\the\figno}%
\writedef{#1\leftbracket fig.\noexpand~\the\figno}%
\ifnum\figno=1\immediate\openout\ffile=figs.tmp\fi\chardef\wfile=\ffile%
\immediate\write\ffile{\noexpand\medskip\noexpand\item{Fig.\ \the\figno. }
\reflabeL{#1\hskip.55in}\pctsign}\global\advance\figno by1\findarg}
\newwrite\tfile\global\newcount\tabno \global\tabno=1
\def\tab{tab.~\the\tabno\ntab}
\def\ntab#1{\xdef#1{Table~\the\tabno}%
\writedef{#1\leftbracket tab.\noexpand~\the\tabno}%
\ifnum\tabno=1\immediate\openout\tfile=tabs.tmp\fi\chardef\wfile=\tfile%
\immediate\write\tfile{\noexpand\medskip\noexpand\item{Table\ \the\tabno. }
\reflabeL{#1\hskip.55in}\pctsign}\global\advance\tabno by1\findarg}
\def\xfig{\expandafter\xf@g}\def\xf@g fig.\penalty\@M\ {}
\def\figs#1{figs.~\f@gs #1{\hbox{}}}
\def\f@gs#1{\edef\next{#1}\ifx\next\em@rk\def\next{}\else
\ifx\next#1\xfig #1\else#1\fi\let\next=\f@gs\fi\next}
\newwrite\lfile
{\escapechar-1\xdef\pctsign{\string\%}\xdef\leftbracket{\string\{}
\xdef\rightbracket{\string\}}\xdef\numbersign{\string\#}}

\def\writestop{\def\writestoppt{\immediate\write\lfile{\string\pageno%
\the\pageno\string\startrefs\leftbracket\the\refno\rightbracket%
\string\def\string\secsym\leftbracket\secsym\rightbracket%
\string\secno\the\secno\string\meqno\the\meqno}\immediate\closeout\lfile}}
\def\writestoppt{}\def\writedef#1{}
\def\seclab#1{\xdef #1{\the\secno}\writedef{#1\leftbracket#1}\wrlabeL{#1=#1}}
\def\subseclab#1{\xdef #1{\secsym\the\subsecno}%
\writedef{#1\leftbracket#1}\wrlabeL{#1=#1}}
\newwrite\tfile \def\writetoca#1{}
\def\leaderfill{\leaders\hbox to 1em{\hss.\hss}\hfill}
\def\writetoc{\immediate\openout\tfile=toc.tmp
   \def\writetoca##1{{\edef\next{\write\tfile{\noindent ##1
   \string\leaderfill {\noexpand\number\pageno} \par}}\next}}}

%
\catcode`\@=12 
%
\edef\tfontsize{\ifx\answ\bigans scaled\magstep3\else scaled\magstep4\fi}
\font\titlerm=cmr10 \tfontsize \font\titlerms=cmr7 \tfontsize
\font\titlermss=cmr5 \tfontsize \font\titlei=cmmi10 \tfontsize
\font\titleis=cmmi7 \tfontsize \font\titleiss=cmmi5 \tfontsize
\font\titlesy=cmsy10 \tfontsize \font\titlesys=cmsy7 \tfontsize
\font\titlesyss=cmsy5 \tfontsize \font\titleit=cmti10 \tfontsize
\skewchar\titlei='177 \skewchar\titleis='177 \skewchar\titleiss='177
\skewchar\titlesy='60 \skewchar\titlesys='60 \skewchar\titlesyss='60
\def\titlefont{\def\rm{\fam0\titlerm}
\textfont0=\titlerm \scriptfont0=\titlerms \scriptscriptfont0=\titlermss
\textfont1=\titlei \scriptfont1=\titleis \scriptscriptfont1=\titleiss
\textfont2=\titlesy \scriptfont2=\titlesys \scriptscriptfont2=\titlesyss
\textfont\itfam=\titleit \def\it{\fam\itfam\titleit}\rm}
 \ifx\answ\bigans\else scaled\magstep1\fi
\ifx\answ\bigans\def\abstractfont{\tenpoint}\else
\font\abssl=cmsl10 scaled \magstep1
\font\absrm=cmr10 scaled\magstep1 \font\absrms=cmr7 scaled\magstep1
\font\absrmss=cmr5 scaled\magstep1 \font\absi=cmmi10 scaled\magstep1
\font\absis=cmmi7 scaled\magstep1 \font\absiss=cmmi5 scaled\magstep1
\font\abssy=cmsy10 scaled\magstep1 \font\abssys=cmsy7 scaled\magstep1
\font\abssyss=cmsy5 scaled\magstep1 \font\absbf=cmbx10 scaled\magstep1
\skewchar\absi='177 \skewchar\absis='177 \skewchar\absiss='177
\skewchar\abssy='60 \skewchar\abssys='60 \skewchar\abssyss='60
\def\abstractfont{\def\rm{\fam0\absrm}
\textfont0=\absrm \scriptfont0=\absrms \scriptscriptfont0=\absrmss
\textfont1=\absi \scriptfont1=\absis \scriptscriptfont1=\absiss
\textfont2=\abssy \scriptfont2=\abssys \scriptscriptfont2=\abssyss
\textfont\itfam=\bigit \def\it{\fam\itfam\bigit}\def\footnotefont{\tenpoint}%
\textfont\slfam=\abssl \def\sl{\fam\slfam\abssl}%
\textfont\bffam=\absbf \def\bf{\fam\bffam\absbf}\rm}\fi
\def\tenpoint{\def\rm{\fam0\tenrm}
\textfont0=\tenrm \scriptfont0=\sevenrm \scriptscriptfont0=\fiverm
\textfont1=\teni  \scriptfont1=\seveni  \scriptscriptfont1=\fivei
\textfont2=\tensy \scriptfont2=\sevensy \scriptscriptfont2=\fivesy
\textfont\itfam=\tenit
\def\it{\fam\itfam\tenit}\def\footnotefont{\ninepoint}%
\textfont\bffam=\tenbf \def\bf{\fam\bffam\tenbf}\def\sl{\fam\slfam\tensl}\rm}
\font\ninerm=cmr9 \font\sixrm=cmr6 \font\ninei=cmmi9 \font\sixi=cmmi6
\font\ninesy=cmsy9 \font\sixsy=cmsy6 \font\ninebf=cmbx9
\font\nineit=cmti9 \font\ninesl=cmsl9 \skewchar\ninei='177
\skewchar\sixi='177 \skewchar\ninesy='60 \skewchar\sixsy='60
\def\ninepoint{\def\rm{\fam0\ninerm}
\textfont0=\ninerm \scriptfont0=\sixrm \scriptscriptfont0=\fiverm
\textfont1=\ninei \scriptfont1=\sixi \scriptscriptfont1=\fivei
\textfont2=\ninesy \scriptfont2=\sixsy \scriptscriptfont2=\fivesy
\textfont\itfam=\ninei \def\it{\fam\itfam\nineit}\def\sl{\fam\slfam\ninesl}%
\textfont\bffam=\ninebf \def\bf{\fam\bffam\ninebf}\rm}
%
%

\hyphenation{anom-aly anom-alies coun-ter-term coun-ter-terms}
\def\inv{^{\raise.15ex\hbox{${\scriptscriptstyle -}$}\kern-.05em 1}}

\def\Dsl{\,\raise.15ex\hbox{/}\mkern-13.5mu D} 
\def\dsl{\raise.15ex\hbox{/}\kern-.57em\partial}

\font\bigit=cmti10 scaled \magstep1
\def\lspace{\ifx\answ\bigans{}\else\qquad\fi}
\def\lbspace{\ifx\answ\bigans{}\else\hskip-.2in\fi} 
\def\boxeqn#1{\vcenter{\vbox{\hrule\hbox{\vrule\kern3pt\vbox{\kern3pt
           \hbox{${\displaystyle #1}$}\kern3pt}\kern3pt\vrule}\hrule}}}
\def\mbox#1#2{\vcenter{\hrule \hbox{\vrule height#2in
               \kern#1in \vrule} \hrule}}  
%

\def\e#1{{\rm e}^{^{\textstyle#1}}}

\def\om#1#2{\omega^{#1}{}_{#2}}

\def\darr#1{\raise1.5ex\hbox{$\leftrightarrow$}\mkern-16.5mu #1}

\def\roughly#1{\raise.3ex\hbox{$#1$\kern-.75em\lower1ex\hbox{$\sim$}}}

\input epsf.tex


\def\IB{\relax\hbox{$\inbar\kern-.3em{\rm B}$}}
\def\IC{\relax\hbox{$\inbar\kern-.3em{\rm C}$}}
\def\ID{\relax\hbox{$\inbar\kern-.3em{\rm D}$}}
\def\IE{\relax\hbox{$\inbar\kern-.3em{\rm E}$}}
\def\IF{\relax\hbox{$\inbar\kern-.3em{\rm F}$}}
\def\IG{\relax\hbox{$\inbar\kern-.3em{\rm G}$}}
\def\IGa{\relax\hbox{${\rm I}\kern-.18em\Gamma$}}
\def\IH{\relax{\rm I\kern-.18em H}}
\def\IK{\relax{\rm I\kern-.18em K}}
\def\II{\relax{\rm I\kern-.18em I}}
\def\IL{\relax{\rm I\kern-.18em L}}
\def\IP{\relax{\rm I\kern-.18em P}}
\def\IR{\relax{\rm I\kern-.18em R}}
\def\IZ{\relax\ifmmode\mathchoice {\hbox{\cmss Z\kern-.4em Z}}{\hbox{\cmss
Z\kern-.4em Z}} {\lower.9pt\hbox{\cmsss Z\kern-.4em Z}}
{\lower1.2pt\hbox{\cmsss Z\kern-.4em Z}}\else{\cmss Z\kern-.4em Z}\fi}

\def\IB{\relax{\rm I\kern-.18em B}}
\def\IC{{\relax\hbox{$\inbar\kern-.3em{\rm C}$}}}
\def\ID{\relax{\rm I\kern-.18em D}}
\def\IE{\relax{\rm I\kern-.18em E}}
\def\IF{\relax{\rm I\kern-.18em F}}


\def\CW {{\cal W}}

\def\p{\partial}





\def\demi{{1\over 2}}


\def\f{\phi}

\def\a{\alpha}
\def\b{\beta}
  
\def\d{\delta}  \def\D{\Delta}
\def\m{\mu}

\def\r{\rho}
\def\l{\lambda} \def\L{\Lambda}

\def\e{\epsilon}

\def\|{\Big|}
\def\({\Big(}   \def\){\Big)}
\def\[{\Big[}   \def\]{\Big]}



\def\paper#1#2#3#4{#1, {\sl #2}, #3 {\tt #4}}

\def\hh{hep-th/}


\def\PLB#1#2#3{Phys. Lett.~{\bf B#1} (#2) #3}
\def\NPB#1#2#3{Nucl. Phys.~{\bf B#1} (#2) #3}
\def\PRL#1#2#3{Phys. Rev. Lett.~{\bf #1} (#2) #3}
\def\CMP#1#2#3{Comm. Math. Phys.~{\bf #1} (#2) #3}
\def\PRD#1#2#3{Phys. Rev.~{\bf D#1} (#2) #3}
\def\MPL#1#2#3{Mod. Phys. Lett.~{\bf #1} (#2) #3}
\def\IJMP#1#2#3{Int. Jour. Mod. Phys.~{\bf #1} (#2) #3}


\def\unlockat{\catcode`\@=11}
\def\lockat{\catcode`\@=12}

\unlockat


\def\newsec#1{\global\advance\secno by1\message{(\the\secno. #1)}
\global\subsecno=0\global\subsubsecno=0\eqnres@t\noindent {\bf\the\secno. #1}
\writetoca{{\secsym} {#1}}\par\nobreak\medskip\nobreak}
\global\newcount\subsecno \global\subsecno=0
\def\subsec#1{\global\advance\subsecno by1\message{(\secsym\the\subsecno.
#1)}
\ifnum\lastpenalty>9000\else\bigbreak\fi\global\subsubsecno=0
\noindent{\it\secsym\the\subsecno. #1}
\writetoca{\string\quad {\secsym\the\subsecno.} {#1}}
\par\nobreak\medskip\nobreak}
\global\newcount\subsubsecno \global\subsubsecno=0
\def\subsubsec#1{\global\advance\subsubsecno by1
\message{(\secsym\the\subsecno.\the\subsubsecno. #1)}
\ifnum\lastpenalty>9000\else\bigbreak\fi
\noindent\quad{\secsym\the\subsecno.\the\subsubsecno.}{#1}
\writetoca{\string\qquad{\secsym\the\subsecno.\the\subsubsecno.}{#1}}
\par\nobreak\medskip\nobreak}

\def\subsubseclab#1{\DefWarn#1\xdef #1{\noexpand\hyperref{}{subsubsection}%
{\secsym\the\subsecno.\the\subsubsecno}%
{\secsym\the\subsecno.\the\subsubsecno}}%
\writedef{#1\leftbracket#1}\wrlabeL{#1=#1}}
\lockat

\def\dbend{\lower3.5pt\hbox{\manual\char127}}


\def\boxit#1{\vbox{\hrule\hbox{\vrule\kern8pt
\vbox{\hbox{\kern8pt}\hbox{\vbox{#1}}\hbox{\kern8pt}}
\kern8pt\vrule}\hrule}}

\def\mathboxit#1{\vbox{\hrule\hbox{\vrule\kern8pt\vbox{\kern8pt
\hbox{$\displaystyle #1$}\kern8pt}\kern8pt\vrule}\hrule}}


\def\inbar{\,\vrule height1.5ex width.4pt depth0pt}

\font\cmss=cmss10 \font\cmsss=cmss10 at 7pt


\lref\simons{ J. Cheeger and J. Simons, {\it Differential Characters and
Geometric Invariants},  Stony Brook Preprint, (1973), unpublished.}

\lref\cargese{ L.~Baulieu, {\it Algebraic quantization of gauge theories},
Perspectives in fields and particles, Plenum Press, eds. Basdevant-Levy,
Cargese Lectures 1983}

\lref\antifields{ L. Baulieu, M. Bellon, S. Ouvry, C.Wallet, Phys. Lett.
B252 (1990) 387; M.  Bocchichio, Phys. Lett. B187 (1987) 322;  Phys. Lett.
B192 (1987) 31; R.  Thorn    Nucl. Phys.   B257 (1987) 61. }

\lref\thompson{ George Thompson,  Annals Phys. 205 (1991) 130; J. M. F.
Labastida, M. Pernici, Phys. Lett. 212B  (1988) 56; D. Birmingham, M. Blau,
M. Rakowski and G. Thompson, Phys. Rept. 209 (1991) 129.}

\lref\tonin{ Tonin}

\lref\wittensix{ E.  Witten, {\it New  Gauge  Theories In Six Dimensions},
\hh{9710065}. }

\lref\orlando{ O. Alvarez, L. A. Ferreira and J. Sanchez Guillen, {\it  A New
Approach to Integrable Theories in any Dimension}, hep-th/9710147.}

\lref\wittentopo{ E.  Witten,  {\it  Topological Quantum Field Theory},
\hh9403195, Commun.  Math. Phys.  {117} (1988)353.  }

\lref\wittentwist{ E.  Witten, {\it Supersymmetric Yang--Mills theory on a
four-manifold}, J.  Math.  Phys.  {35} (1994) 5101.}

\lref\west{ L.~Baulieu, P.~West, {\it Six Dimensional TQFTs and  Self-dual
Two-Forms,} Phys. Lett. B{\bf 436} (1998) 97, /hep-th/9805200}

\lref\bv{ I. A. Batalin and V. A. Vilkowisky,    Phys. Rev.   D28  (1983)
2567\semi M. Henneaux,  Phys. Rep.  126   (1985) 1\semi M. Henneaux and C.
Teitelboim, {\it Quantization of Gauge Systems}
  Princeton University Press,  Princeton (1992).}

\lref\kyoto{ L. Baulieu, E. Bergschoeff and E. Sezgin, Nucl. Phys.
B307(1988)348\semi L. Baulieu,   {\it Field Antifield Duality, p-Form Gauge
Fields
   and Topological Quantum Field Theories}, hep-th/9512026,
   Nucl. Phys. B478 (1996) 431.  }

\lref\sourlas{ G. Parisi and N. Sourlas, {\it Random Magnetic Fields,
Supersymmetry and Negative Dimensions}, Phys. Rev. Lett.  43 (1979) 744;
Nucl.  Phys.  B206 (1982) 321.  }

\lref\SalamSezgin{ A.  Salam  and  E.  Sezgin, {\it Supergravities in
diverse dimensions}, vol.  1, p. 119\semi P.  Howe, G.  Sierra and P.
Townsend, Nucl. Phys. B221 (1983) 331.}

\lref\nekrasov{ A. Losev, G. Moore, N. Nekrasov, S. Shatashvili, {\it
Four-Dimensional Avatars of Two-Dimensional RCFT},  hep-th/9509151, Nucl.
Phys.  Proc.  Suppl.   46 (1996) 130\semi L.  Baulieu, A.  Losev,
N.~Nekrasov  {\it Chern-Simons and Twisted Supersymmetry in Higher
Dimensions},  hep-th/9707174, to appear in Nucl.  Phys.  B.  }

\lref\WitDonagi{R.~ Donagi, E.~ Witten, ``Supersymmetric Yang--Mills Theory
and Integrable Systems'', hep-th/9510101, Nucl. Phys. {\bf B}460 (1996)
299-334}
\lref\Witfeb{E.~ Witten, ``Supersymmetric Yang--Mills Theory On A
Four-Manifold,''  hep-th/9403195; J. Math. Phys. {\bf 35} (1994) 5101.}
\lref\Witgrav{E.~ Witten, ``Topological Gravity'', Phys. Lett. 206B:601, 1988}
\lref\witaffl{I. ~ Affleck, J.A.~ Harvey and E.~ Witten,
        ``Instantons and (Super)Symmetry Breaking
        in $2+1$ Dimensions'', Nucl. Phys. {\bf B}206 (1982) 413}
\lref\wittabl{E.~ Witten,  ``On $S$-Duality in Abelian Gauge Theory,''
hep-th/9505186; Selecta Mathematica {\bf 1} (1995) 383}
\lref\wittgr{E.~ Witten, ``The Verlinde Algebra And The Cohomology Of The
Grassmannian'',  hep-th/9312104}
\lref\wittenwzw{E. Witten, ``Non Abelian bosonization in two dimensions,''
Commun. Math. Phys. {\bf 92} (1984)455 }
\lref\witgrsm{E. Witten, ``Quantum field theory, grassmannians and algebraic
curves,'' Commun. Math. Phys. 113:529,1988}
\lref\wittjones{E. Witten, ``Quantum field theory and the Jones
polynomial,'' Commun.  Math. Phys., 121 (1989) 351. }
\lref\witttft{E.~ Witten, ``Topological Quantum Field Theory", Commun. Math.
Phys. {\bf 117} (1988) 353.}
\lref\wittmon{E.~ Witten, ``Monopoles and Four-Manifolds'', hep-th/9411102}
\lref\Witdgt{ E.~ Witten, ``On Quantum gauge theories in two dimensions,''
Commun. Math. Phys. {\bf  141}  (1991) 153}
\lref\witrevis{E.~ Witten,
 ``Two-dimensional gauge theories revisited'', hep-th/9204083; J. Geom.
Phys. 9 (1992) 303-368}
\lref\Witgenus{E.~ Witten, ``Elliptic Genera and Quantum Field Theory'',
Comm. Math. Phys. 109(1987) 525. }
\lref\OldZT{E. Witten, ``New Issues in Manifolds of $SU(3)$ Holonomy,'' {\it
Nucl. Phys.} {\bf B268} (1986) 79 \semi J. Distler and B. Greene, ``Aspects
of (2,0) String Compactifications,'' {\it Nucl. Phys.} {\bf B304} (1988) 1
\semi B. Greene, ``Superconformal Compactifications in Weighted Projective
Space,'' {\it Comm. Math. Phys.} {\bf 130} (1990) 335.}
\lref\bagger{E.~ Witten and J. Bagger, Phys. Lett. {\bf 115B}(1982) 202}
\lref\witcurrent{E.~ Witten,``Global Aspects of Current Algebra'',
Nucl. Phys. B223 (1983) 422\semi ``Current Algebra, Baryons and Quark
Confinement'', Nucl. Phys. B223 (1993) 433}
\lref\Wittreiman{S. B. Treiman, E. Witten, R. Jackiw, B. Zumino, ``Current
Algebra and Anomalies'', Singapore, Singapore: World Scientific ( 1985) }
\lref\Witgravanom{L. Alvarez-Gaume, E.~ Witten, ``Gravitational Anomalies'',
Nucl. Phys. B234:269,1984. }

\lref\nicolai{\paper {H.~Nicolai}{New Linear Systems for 2D Poincar\'e
Supergravities}{\NPB{414}{1994}{299},}{\hh 9309052}.}



\lref\bg{\paper {L.~Baulieu, B.~Grossman}{Monopoles and Topological Field
Theory}{\PLB{214}{1988}{223}.}{}}

\lref\seibergsix{\paper {N.~Seiberg}{Non-trivial Fixed Points of The
Renormalization Group in Six
 Dimensions}{\PLB{390}{1997}{169}}{\hh 9609161}\semi
\paper {O. J.~Ganor, D. R.~Morrison, N.~Seiberg}{
  Branes, Calabi-Yau Spaces, and Toroidal Compactification of the N=1
  Six-Dimensional $E_8$ Theory}{\NPB{487}{1997}{93}}{\hh 9610251}\semi
\paper {O.~Aharony, M.~Berkooz, N.~Seiberg}{Light-Cone
  Description of (2,0) Superconformal Theories in Six
  Dimensions}{Adv. Theor. Math. Phys. {\bf 2} (1998) 119}{\hh 9712117.}}

\lref\cs{\paper {L.~Baulieu}{Chern-Simons Three-Dimensional and
Yang--Mills-Higgs Two-Dimensional Systems as Four-Dimensional Topological
Quantum Field Theories}{\PLB{232}{1989}{473}.}{}}

\lref\beltrami{\paper {L.~Baulieu, M.~Bellon}{Beltrami Parametrization and
String Theory}{\PLB{196}{1987}{142}}{}\semi
\paper {L.~Baulieu, M.~Bellon, R.~Grimm}{Beltrami Parametrization For
Superstrings}{\PLB{198}{1987}{343}}{}\semi
\paper {R.~Grimm}{Left-Right Decomposition of Two-Dimensional Superspace
Geometry and Its BRS Structure}{Annals Phys. {\bf 200} (1990) 49.}{}}

\lref\bbg{\paper {L.~Baulieu, M.~Bellon, R.~Grimm}{Left-Right Asymmetric
Conformal Anomalies}{\PLB{228}{1989}{325}.}{}}

\lref\bonora{\paper {G.~Bonelli, L.~Bonora, F.~Nesti}{String Interactions
from Matrix String Theory}{\NPB{538}{1999}{100},}{\hh 9807232}\semi
\paper {G.~Bonelli, L.~Bonora, F.~Nesti, A.~Tomasiello}{Matrix String Theory
and its Moduli Space}{}{\hh 9901093.}}

\lref\corrigan{\paper {E.~Corrigan, C.~Devchand, D.B.~Fairlie,
J.~Nuyts}{First Order Equations for Gauge Fields in Spaces of Dimension
Greater Than Four}{\NPB{214}{452}{1983}.}{}}

\lref\acha{\paper {B. S.~Acharya, M.~O'Loughlin, B.~Spence}{Higher
Dimensional Analogues of Donaldson-Witten Theory}{\NPB{503}{1997}{657},}{\hh
9705138}\semi
\paper {B. S.~Acharya, J. M.~Figueroa-O'Farrill, M.~O'Loughlin,
B.~Spence}{Euclidean
  D-branes and Higher-Dimensional Gauge   Theory}{\NPB{514}{1998}{583},}{\hh
  9707118.}}

\lref\Witr{\paper{E.~Witten}{Introduction to Cohomological Field   Theories}
{Lectures at Workshop on Topological Methods in Physics (Trieste, Italy, Jun
11-25, 1990), \IJMP{A6}{1991}{2775}.}{}}

\lref\ohta{\paper {L.~Baulieu, N.~Ohta}{Worldsheets with Extended
Supersymmetry} {\PLB{391}{1997}{295},}{\hh 9609207}.}

\lref\gravity{\paper {L.~Baulieu}{Transmutation of Pure 2-D Supergravity
Into Topological 2-D Gravity and Other Conformal Theories}
{\PLB{288}{1992}{59},}{\hh 9206019.}}

\lref\wgravity{\paper {L.~Baulieu, M.~Bellon, R.~Grimm}{Some Remarks on  the
Gauging of the Virasoro and   $w_{1+\infty}$
Algebras}{\PLB{260}{1991}{63}.}{}}

\lref\fourd{\paper {E.~Witten}{Topological Quantum Field
Theory}{\CMP{117}{1988}{353}}{}\semi
\paper {L.~Baulieu, I.M.~Singer}{Topological Yang--Mills Symmetry}{Nucl.
Phys. Proc. Suppl. {\bf 15B} (1988) 12.}{}}

\lref\topo{\paper {L.~Baulieu}{On the Symmetries of Topological Quantum Field
Theories}{\IJMP{A10}{1995}{4483},}{\hh 9504015}\semi
\paper {R.~Dijkgraaf, G.~Moore}{Balanced Topological Field
Theories}{\CMP{185}{1997}{411},}{\hh 9608169.}}

\lref\wwgravity{\paper {I.~Bakas} {The Large $N$ Limit   of Extended
Conformal Symmetries}{\PLB{228}{1989}{57}.}{}}

\lref\wwwgravity{\paper {C. M.~Hull}{Lectures on $\CW$-Gravity,
$\CW$-Geometry and
$\CW$-Strings}{}{\hh 9302110}, and~references therein.}

\lref\wvgravity{\paper {A.~Bilal, V.~Fock, I.~Kogan}{On the origin of
$W$-algebras}{\NPB{359}{1991}{635}.}{}}

\lref\surprises{\paper {E.~Witten} {Surprises with Topological Field
Theories} {Lectures given at ``Strings 90'', Texas A\&M, 1990,}{Preprint
IASSNS-HEP-90/37.}}

\lref\stringsone{\paper {L.~Baulieu, M. B.~Green, E.~Rabinovici}{A Unifying
Topological Action for Heterotic and  Type II Superstring  Theories}
{\PLB{386}{1996}{91},}{\hh 9606080.}}

\lref\stringsN{\paper {L.~Baulieu, M. B.~Green, E.~Rabinovici}{Superstrings
from   Theories with $N>1$ World Sheet Supersymmetry}
{\NPB{498}{1997}{119},}{\hh 9611136.}}

\lref\bks{\paper {L.~Baulieu, H.~Kanno, I.~Singer}{Special Quantum Field
Theories in Eight and Other Dimensions}{\CMP{194}{1998}{149},}{\hh
9704167}\semi
\paper {L.~Baulieu, H.~Kanno, I.~Singer}{Cohomological Yang--Mills Theory
  in Eight Dimensions}{ Talk given at APCTP Winter School on Dualities in
String Theory (Sokcho, Korea, February 24-28, 1997),} {\hh 9705127.}}

\lref\witdyn{\paper {P.~Townsend}{The eleven dimensional supermembrane
revisited}{\PLB{350}{1995}{184},}{\hh9501068}\semi
\paper{E.~Witten}{String Theory Dynamics in Various Dimensions}
{\NPB{443}{1995}{85},}{\hh 9503124}.}

\lref\bfss{\paper {T.~Banks, W. Fischler, S. H.~Shenker,
L.~Susskind}{$M$-Theory as a Matrix Model~:
A~Conjecture}{\PRD{55}{1997}{5112},} {\hh9610043.}}

\lref\seiberg{\paper {N.~Seiberg}{Why is the Matrix Model
Correct?}{\PRL{79}{1997}{3577},} {\hh 9710009.}}

\lref\sen{\paper {A.~Sen}{$D0$ Branes on $T^n$ and Matrix Theory}{Adv.
Theor. Math. Phys.~{\bf 2} (1998) 51,} {\hh 9709220.}}

\lref\laroche{\paper {L.~Baulieu, C.~Laroche} {On Generalized Self-Duality
Equations Towards Supersymmetric   Quantum Field Theories Of
Forms}{\MPL{A13}{1998}{1115},}{\hh  9801014.}}

\lref\bsv{\paper {M.~Bershadsky, V.~Sadov, C.~Vafa} {$D$-Branes and
Topological Field Theories}{\NPB{463} {1996}{420},}{\hh9511222.}}

\lref\vafapuzz{\paper {C.~Vafa}{Puzzles at Large N}{}{\hh 9804172.}}

\lref\dvv{\paper {R.~Dijkgraaf, E.~Verlinde, H.~Verlinde} {Matrix String
Theory}{\NPB{500}{1997}{43},} {\hh9703030.}}

\lref\wynter{\paper {T.~Wynter}{Gauge Fields and Interactions in Matrix
String Theory}{\PLB{415}{1997}{349},}{\hh9709029.}}

\lref\kvh{\paper {I.~Kostov, P.~Vanhove}{Matrix String Partition
Functions}{}{\hh9809130.}}

\lref\ikkt{\paper {N.~Ishibashi, H.~Kawai, Y.~Kitazawa, A.~Tsuchiya} {A
Large $N$ Reduced Model as Superstring}{\NPB{498} {1997}{467},}{\hh
9612115.}}

\lref\ss{\paper {S.~Sethi, M.~Stern} {$D$-Brane Bound States
Redux}{\CMP{194}{1998} {675},}{\hh 9705046.}}

\lref\mns{\paper {G.~Moore, N.~Nekrasov, S.~Shatashvili} {$D$-particle Bound
States and Generalized Instantons}{} {\hh 9803265.}}

\lref\bsh{\paper {L.~Baulieu, S.~Shatashvili} {Duality from Topological
Symmetry}{} {\hh 9811198.}}

\lref\pawu{ {G.~Parisi, Y.S.~Wu} {}{ Sci. Sinica  {\bf 24} {(1981)} {484}.}}

\lref\coulomb{ {L.~Baulieu, D.~Zwanziger, }   {\it Renormalizable Non-Covariant
Gauges and Coulomb Gauge Limit}, {Nucl. Phys. B {\bf 548 } (1999) 527,} {\hh
9807024}.}

\lref\danzinn{  {J.~Zinn-Justin, D.~Zwanziger, } {}{Nucl. Phys. B  {\bf
295} (1988) {297}.}{}}

\lref\danlau{ {L.~Baulieu, D.~Zwanziger, } {\it Equivalence of Stochastic
Quantization and the Faddeev-Popov Ansatz,
  }{Nucl. Phys. B  {\bf 193 } (1981) {163}.}{}}

\lref\munoz{ { A.~Munoz Sudupe, R. F. Alvarez-Estrada, } {}
Phys. Lett. {\bf 164} (1985) 102; {} {\bf 166B} (1986) 186. }

\lref\okano{ { K.~Okano, } {}
Nucl. Phys. {\bf B289} (1987) 109; {} Prog. Theor. Phys.
suppl. {\bf 111} (1993) 203. }

\lref\singer{
 I. M. Singer, { Comm. Math. Phys. {\bf 60} (1978) 7.}}

\lref\neu{ {H.~Neuberger,} {Phys. Lett. B {\bf 295}
(1987) {337}.}{}}

\lref\testa{ {M.~Testa,} {}{Phys. Lett. B {\bf 429}
(1998) {349}.}{}}

\lref\Martin{ L.~Baulieu and M. Schaden, {\it Gauge Group TQFT and Improved
Perturbative Yang--Mills Theory}, {  Int. Jour. Mod.  Phys. A {\bf  13}
(1998) 985},   hep-th/9601039.}

\lref\baugros{ {L.~Baulieu, B.~Grossman, } {\it A topological Interpretation
of  Stochastic Quantization} {Phys. Lett. B {\bf  212} {(1988)} {351}.}}

\lref\bautop{ {L.~Baulieu}{ \it Stochastic and Topological Field Theories},
{Phys. Lett. B {\bf   232} (1989) {479}}{}; {}{ \it Topological Field Theories
And Gauge Invariance in Stochastic Quantization}, {Int. Jour. Mod.  Phys. A
{\bf  6} (1991) {2793}.}{}}

\lref\samson{ {L.~Baulieu, S. L.~Shatashvili, { \it Duality from Topological
Symmetry}, {JHEP {\bf 9903} (1999) 011, hep-th/9811198.}}}{}

\lref\halpern{ {H. S.~Chan, M. B.~Halpern}{}, {Phys. Rev. D {\bf   33} (1986)
{540}.}}

\lref\yue{ {Yue-Yu}, {Phys. Rev. D {\bf   33} (1989) {540}.}}

\lref\neuberger{ {H.~Neuberger,} {\it Non-perturbative gauge Invariance},
{ Phys. Lett. B {\bf 175} (1986) {69}.}{}}

\lref\huffel{ {P. H.~Daamgard, H. Huffel},  {}{Phys. Rep. {\bf 152} (1987)
{227}.}{}}

\lref\creutz{ {M.~Creutz},  {\it Quarks, Gluons and  Lattices, }  Cambridge
University Press 1983, pp 101-107.}

\lref\zinn{ {J. Zinn-Justin, }  {Nucl. Phys. B {\bf  275} (1986) {135}.}}

\lref\shamir{  {Y.~Shamir,  } {\it Lattice Chiral Fermions
  }{ Nucl.  Phys.  Proc.  Suppl.  {\bf } 47 (1996) 212,  hep-lat/9509023;
V.~Furman, Y.~Shamir, Nucl. Phys. B {\bf 439 } (1995), hep-lat/9405004.}}

 \lref\kaplan{ {D. B.~Kaplan, }  {\it A Method for Simulating Chiral
Fermions on the Lattice,} Phys. Lett. B {\bf 288} (1992) 342; {\it Chiral
Fermions on the Lattice,}  {  Nucl. Phys. B, Proc. Suppl.  {\bf 30} (1993)
597.}}

\lref\ZZ{ {I. Zahed, D. Zwanziger, } {\it  Zero Color Magnetization in
QCD Matter, }{Phys. Rev. D61 (2000) 037501.}}

\lref\neubergerr{ {H.~Neuberger, } {\it Chirality on the Lattice},
hep-lat/9808036.}


\lref\nielsen{N. K. Nielsen,
               {\it On the gauge dependence of spontaneous
                    symmetry breaking in gauge theories},
               Nucl.\ Phys.\ {\bf B101} (1975) 173.}

\lref\cutkosky{R. E. Cutkosky,
               {\it The Gribov horizon},
               Phys.\ Rev.\ {\bf D30} (1984) 447.}

\lref\paretal{C. Parrinello, S. Petrarca and A. Vladikas,
               {\it A preliminary study of the Gribov ambiguity
                    in lattice $SU(3)$ Coulomb gauge},
               Phys.\ Lett.\ {\bf B268} (1991) 236.}

\lref\hybrid{K. M. Decker and Ph. de Forcrand,
               {\it Pure $SU(2)$ lattice gauge theory on
                    $32^4$ lattices},
               Nucl.\ Phys.\ {\bf B} (Proc. Suppl.) {\bf 17} (1990) 567.}

\lref\gaugefix{A. Cucchieri and T. Mendes,
               {\it Critical slowing-down in $SU(2)$ Landau
                    gauge-fixing algorithms},
               Nucl.\ Phys.\ {\bf B471} (1996) 263.}

\lref\gribov{V. N. Gribov,
               {\it Quantization of non-Abelian gauge theories},
               Nucl.\ Phys.\ {\bf B139} (1978) 1.}

\lref\coul{D. Zwanziger,
               {\it Renormalization in the Coulomb gauge and
                    order parameter for confinement in QCD},
               Nucl.\ Phys.\ {\bf  B518} (1998) 237.}

\lref\christlee{N. H. Christ and T. D. Lee,
               {\it Operator ordering and Feynman rules in gauge theories},
               Phys.\ Rev.\  {\bf D22} (1980) 939.}

\lref\latcoulh{D. Zwanziger,
               {\it Lattice Coulomb Hamiltonian and static
                    color-Coulomb field},
               Nucl.\ Phys.\ {\bf B485} (1997) 185.}

\lref\critical{D. Zwanziger,
               {\it Critical limit of lattice gauge theory},
               Nucl.\ Phys.\ {\bf B378} (1992) 525.}

\lref\vanish{D. Zwanziger,
               {\it Vanishing of zero-momentum lattice gluon
                    propagator and color confinement},
               Nucl.\ Phys.\ {\bf B364} (1991) 127.}

\lref\cuzwsc{A. Cucchieri and D. Zwanziger,
               {\it Static color-Coulomb force},
               Phys.\ Rev.\ Lett.\ {\bf 78} (1997) 3814.}

\lref\Szcz{A. Szczepaniak et al.,
               {\it Glueball spectroscopy in a relativistic
                    many-body approach to hadron structure},
               Phys.\ Rev.\ Lett.\ {\bf 76} (1996) 2011.}

\lref\Robertson{D. G. Robertson et al.,
               {\it  Renormalized effective QCD Hamiltonian: gluonic sector},
               Phys.\ Rev.\ {\bf D59} (1999) 074019.}

\lref\rgcoul{A. Cucchieri and D. Zwanziger,
               {\it Renormalization-group calculation of color-Coulomb
                potential},
                hep-th/0008248.}

\lref\zbgr{L. Baulieu and D. Zwanziger,
               {\it QCD$_4$ from a five-dimensional point of view},
               Nucl.\ Phys.\ {\bf B581} (2000) 604.}

\lref\bgz{L. Baulieu, P. A. Grassi and D. Zwanziger,
               {\it Gauge and topological symmetries in the bulk
                    quantization of gauge theories},
               hep-th/0006036 and Nucl.\ Phys.\ {\bf B}
						(to be published).}

\lref\fmr{A. Cucchieri,
               {\it Numerical study of the fundamental modular
                    region in the minimal Landau gauge}.
               Nucl.\ Phys.\ {\bf B521} (1998) 365.}

\lref\recoul{L. Baulieu and D. Zwanziger,
               {\it Renormalizable non-covariant gauges and Coulomb
                    gauge limit},
               Nucl.\ Phys.\ {\bf B548} (1999) 527.}

\lref\gluonasym{D. Becirevic et al.,
               {\it Asymptotic scaling of the gluon propagator on
                    the lattice},
               Phys.\ Rev.\ {\bf D61} (2000) 114508.}

\lref\stringtens{J. Fingberg, U. Heller and F. Karsch,
               {\it Scaling and asymptotic scaling in the $SU(2)$
                    gauge theory},
               Nucl.\ Phys.\ {\bf B392} (1993) 493.}

\lref\bps{C. Bernard, C. Parrinello and A. Soni,
               {\it A lattice study of the gluon propagator
                    in momentum space},
               Phys.\ Rev.\ {\bf D49} (1994) 1585.}

\lref\gluonadelaide{D. B. Leinweber et al.,
               {\it Asymptotic scaling and infrared behavior
                    of the gluon propagator},
               Phys.\ Rev.\ {\bf D60} (1999) 094507,
             Erratum-ibid.\ {\bf D61} (2000) 079901.}

\lref\discreti{A. Cucchieri and T. Mendes,
               {\it Gauge fixing and gluon propagator in
                    $\lambda$-gauges},
               in ``Strong and Electroweak Matter '98'',
               edited by J.\ Ambj\o rn et al.
               (World Scientific, Singapore, 1999);
               A. Cucchieri,
               {\it The lattice gluon propagator into
                    the next millennium},
               in ``Understanding Deconfinement in QCD'',
               edited by D.\ Blaschke, F. Karsch and C. D. Roberts
               (World Scientific, Singapore, 2000).}

\lref\gluonalf{B. Alles et al.,
               {\it $\alpha_s$ from the nonperturbatively
                    renormalized lattice three-gluon vertex,}
               Nucl.\ Phys.\ {\bf B502} (1997) 325;
               P. Boucaud et al.,
               {\it Lattice calculation of $\alpha_s$ in
                    momentum scheme},
               JHEP {\bf 10} (1998) 017.}

\lref\gluonhot{A. Cucchieri, F. Karsch and P. Petreczky,
               {\it Magnetic screening in hot non-Abelian gauge
                    theory},
               hep-lat/0004027 and Phys.\ Lett.\ {\bf B}
               (to be published).}

\lref\gnoise{A. Cucchieri,
               {\it Gribov copies in the minimal Landau gauge:
                    the influence on gluon and ghost propagators},
               Nucl.\ Phys.\ {\bf B508} (1997) 353.}


\nfig\compar{Plot of the gluon propagators
$D^{\rm tr}(\vk)$ and $D_{44}(\vk)$ as a function
of the square of the lattice momentum $\vk^2$ for
$L = 28$ (symbols $\ast$ and $\bigcirc$ respectively) and
$L = 30$ (symbols $\triangle$ and $\bigtriangledown$ respectively).
Notice the logarithmic scale in the $y$ axis.
Error bars are one standard deviation.}

\nfig\fitp{Plot of the gluon propagator $D^{\rm tr}(\vk)$
as a function
of the square of the lattice momentum $\vk^2$
for $L = 14 (\ast), \, 20 (\triangle)$ and $26 (\bigcirc)$ and fits of
these data using Eq.\ \mod\ with the parameter values
reported in Table 1.
Error bars are one standard deviation.}

\nfig\fits{Plot of the gluon propagator $D^{\rm tr}(\vk)$
as a function
of the square of the lattice momentum $\vk^2$
for $L = 16 (\ast), \, 22 (\triangle)$ and $28 (\bigcirc)$ and fits of
these data using Eq.\ \mod\ with the parameter values
reported in Table 1.
Error bars are one standard deviation.}

\nfig\fitt{Plot of the gluon propagator $D^{\rm tr}(\vk)$
as a function
of the square of the lattice momentum $\vk^2$
for $L = 18 (\ast), \, 24 (\triangle)$ and $30 (\bigcirc)$ and fits of
these data using Eq.\ \mod\ with the parameter values
reported in Table 1.
Error bars are one standard deviation.}

\nfig\zz{Fit of the parameter $z(L)$
(see Table 1) as a function of $1/L$ using
$a - b/L^c$. We obtain $a = 17.2(249)$, $b = 35.9(117)$,
$c = 0.38(83)$, $\chi^2/$d.o.f.\ $= 0.89$ and
goodness-of-fit $Q = 50.1 \%$.}

\nfig\aa{Fit of the parameter $\alpha(L)$
(see Table 1) as a function of $1/L$ using
$a + \log(1 + b/L)$. We obtain $a = 0.48(5)$, $b = 4.0(13)$,
$\chi^2/$d.o.f.\ $= 0.66$ and goodness-of-fit $Q = 70.6 \%$.}

\nfig\xx{Fit of the parameter $x(L)$
(see Table 1) as a function of $1/L$ using
$a + \log(1 + b/L)$. We obtain $a = -0.2(1)$, $b = 10.0(34)$,
$\chi^2/$d.o.f.\ $= 0.94$ and goodness-of-fit $Q = 47.4 \%$.}

\nfig\yy{Fit of the parameter $y^2(L)$
(see Table 1) as a function of $1/L$ using
$a + \log(1 + b/L)$. We obtain $a = 0.45(1)$, $b = 13.6(45)$,
$\chi^2/$d.o.f.\ $= 1.05$ and goodness-of-fit $Q = 39.3 \%$.}

\nfig\rz{Fit of the product $(rz)(L)$
(see Table 1) as a function of $1/L$ using
$\exp(a)/L^b$. We obtain $a = 5.6(4)$, $b = 1.1(1)$,
$\chi^2/$d.o.f.\ $= 0.36$ and goodness-of-fit $Q = 92.6 \%$.}

\nfig\zero{Fit of the zero-momentum transverse
gluon propagator $D^{\rm tr}(0,L)$ as a function of $1/L$
using $\exp(a)/L^b$. Considering the data for
$L \geq 16$, we
 obtain $a = 2.43(4)$, $b = 0.50(1)$,
$\chi^2/$d.o.f.\ $= 0.69$ and goodness-of-fit $Q = 65.6 \%$.}

\nfig\fitinfall{Plot of the gluon propagator $D^{\rm tr}(\vk)$
as a function
of the square of the lattice momentum $\vk^2$
for $L = 28 (\ast)$ and $30 (\triangle)$ and fits of
these data using Eq.\ \mod\ with the parameter values
and low momentum cut-offs
reported in the fourth (solid line) and fifth
(dotted line) rows of Table 2.
Error bars are one standard deviation.}

\nfig\alphafit{Plot of the running coupling constant
${ {12} \over {11} } g_{\rm coul}^2(\vk) \equiv
\vk^2 D_{44}(\vk)$ as a function
of the square of the lattice momentum $\vk^2$
for $L = 28 (\ast)$ and
$30 (\triangle)$ and fits of
these data using Eq.\ \mrg\ with the parameter values
and low momentum cut-offs
reported in the second row of Table 3.
Error bars are one standard deviation.}

\ntab\tabi{Fit of the transverse gluon propagator
$D^{\rm tr}(\vk)$ using Eq.\ \mod. The resulting fitting
parameters, $\chi^2/$d.o.f.\ and goodness-of-fit $Q$ are
reported for each lattice side $L$.}

\ntab\tabinfi{Extrapolation to infinite lattice side $L$
for the five fitting parameters appearing in Eq.\ \mod\
(we always have $r = 0$). Five different cases are
considered (see Section 2).}

\ntab\tabig{Fit of the running coupling constant
 ${ {12} \over {11} } g_{\rm coul}^2(\vk) \equiv \vk^2 D_{44}(\vk)$
using Eq.\ \mrg. The resulting fitting
parameters, $\chi^2/$d.o.f.\ and goodness-of-fit $Q$ are
reported for three different values of the low-momentum
cut $\vk^2_{\rm min}$.}




\Title{\vbox
{\baselineskip 10pt
\hbox{hep-lat/0008026} \vskip 1mm
\hbox{NYU-TH-PH-19.8.00} \vskip 1mm
\hbox{BI-TP 2000/19}
 \hbox{   }
}}
{\vbox{\vskip -30 true pt
\centerline{ Numerical Study of Gluon Propagator   }
\medskip
 \centerline{ and Confinement Scenario in Minimal Coulomb Gauge  }
\medskip
\vskip4pt }}
\centerline{{\bf Attilio Cucchieri}$^{\star}$\foot{
Address after February 1st, 2001:
IFSC-USP, Caixa Postal 369, 13560-970 S\~ao Carlos, SP, Brazil.}
  and  {\bf  Daniel
Zwanziger}$^{ \ddag}$}
\centerline{attilio@Physik.Uni-Bielefeld.DE, Daniel.Zwanziger@nyu.edu}
\vskip 0.5cm
\centerline{\it $^{\star}$ Fakult\"at f\"ur Physik, Universit\"at Bielefeld,
                 D-33615 Bielefeld, GERMANY}

\centerline{\it $^{\ddag}$   Physics Department, New York University,
New York,  NY 10003,  USA}

\medskip
\vskip  1cm
\noindent

\def\vx{\vec{x}}
\def\vy{\vec{y}}
\def\vk{\vec{k}}

\def\hi{\hat{i}}

We present numerical results in $SU(2)$ lattice gauge theory for the
space-space and
time-time components of the gluon propagator at equal time in the minimal
Coulomb gauge.
It is
found that the equal-time would-be physical 3-dimensionally transverse
gluon propagator
$D^{\rm tr}(\vk)$ vanishes at $\vk = 0$ when extrapolated to infinite
lattice volume,
whereas the instantaneous color-Coulomb potential $D_{44}(\vk)$ is strongly
enhanced at
$\vk = 0$.  This has a natural interpretation in a confinement scenario in
which
the would-be physical gluons leave the physical spectrum while the long-range
Coulomb force confines color.  Gribov's formula
$D^{\rm tr}(\vk) = (|\vk|/2)[(\vk^2)^2 + M^4]^{1/2}$ provides an
excellent fit to our data for the 3-dimensionally transverse equal-time
gluon propagator $D^{\rm tr}(\vk)$ for relevant values of $\vk$.

\Date{\ }

\def\e{\epsilon}
\def\demi{{1\over 2}}

\def\a{\alpha}
\def\b{\beta}
\def\d{\delta}

\def\m{\mu}

\def\om{\omega}
\def\r{\rho}

\def\l{\lambda}
\def\L{\Lambda}

\def\t{\theta}

\newsec{Introduction}

Wilson's lattice gauge theory provides a regularized formulation of gauge
theory that
is manifestly gauge invariant, and numerical simulations do not require
gauge fixing.
However gauge fixing on the lattice is advantageous to gain control of the
critical or
continuum limit, for this makes available the strong  results of
gauge-fixed continuum
renormalization theory.  For example one may prove in continuum
renormalization theory that a certain quantity, such as the running
coupling constant,
defined in terms of gluon correlation functions in some gauge, is finite
when the
cut-off is removed.  Then the corresponding lattice quantity,
defined in the corresponding lattice gauge, should be finite in the
critical limit, and it
becomes of interest to make a numerical determination of that quantity.
Moreover one may determine by numerical fit the location of
the poles of propagators which, according to the Nielsen identities,
is independent of the gauge parameters \nielsen.
Finally, if one
has in hand a confinement scenario in a particular gauge, then it is
possible to test its predictions numerically for gauge-fixed quantities.
Although the scenario
may look quite different in different gauges, nevertheless any one of them
provides a
valid perspective.

	Previous numerical studies of the Coulomb gauge
were reported in \cutkosky\ and \paretal.
We present here a numerical study of the gluon
propagator in $SU(2)$ lattice gauge theory, without quarks,
in the minimal Coulomb gauge
(defined in the Appendix). Simulations have been done at
$\beta = 2.2\, $ for 9 different lattice volumes $V = L^4$, with
$L = 14,\, 16,\, 18,\, 20,\, 22,\, 24,\, 26,\, 28$ and $ 30 $.
(A total of 2420
configurations have been generated, from 50 configurations for
$30^4$ up to 600 for $14^4 $.) The procedure is to first
equilibrate
ungauge-fixed configurations $U$ according to the Wilson action
using a hybrid over-relaxed algorithm \hybrid.
Then statistically independent configurations $U$ are gauge fixed to the
minimal lattice Coulomb gauge by a minimization that is effected
using a stochastic over-relaxation algorithm described in
\gaugefix\ with accuracy
 $\langle (\p_i A_i)^2 \rangle \leq 10^{-16}$, where the average
is taken on each time slice separately and
$\p_i A_i$ is defined in Eq.\ (A.4).  Finally the components
$D^{\rm tr}(\vec{k})$ and $D_{44}(\vec{k})$ of the {\it equal-time} gluon
correlator are evaluated. The lattice Coulomb gauge is more easily
accessible to
numerical study than the Landau gauge because each time-slice contributes
separately to the
numerical average which,  for a lattice of volume $30^4$, gives a factor of
30 gain.
The total computer time devoted to this project so far is about 500 days
on a 500 MHz ALPHA work-station\foot{We thank Jorge L.\ deLyra for kindly
providing us with access to the cluster of ALPHA work-stations
at the Department of Mathematical Physics (DFMA) of the University of
S\~ao Paulo (USP).}.

The results provide a test of the  confinement scenario
that was
originally proposed by Gribov \gribov\
and elaborated in~\coul.  The confinement scenario is particularly
transparent in the Coulomb gauge because it is a physical gauge in the
sense that the constraints are solved
exactly, including Gauss's law
$D_iE_i \equiv \p_i E_i + [A_i, E_i] = \r_{\rm qu}$,
and the Hilbert space has
positive metric.  Here $E_i$ is the color-electric field,
$D_i$ is the gauge-covariant derivative, and $\r_{\rm qu}$ is the
color-charge density of quarks.
Moreover the gauge fixing, described in the Appendix, is done independently
within each time-slice so that the equal-time
Euclidean and Minkowskian of the
correlation functions are identical.

	In the Coulomb gauge, the 3-vector potential $A_i$ is transverse,
$\p_iA_i =
0$, so
$A_i = A_i^{\rm tr}$.  Gauss's law is solved by
\eqn\dgauss{\eqalign{
E_i = E_i^{\rm tr} - \p_i\f ,
}}
where the color-Coulomb field $\f$ is given by
\eqn\gauss{\eqalign{
\f = M^{-1}\r_{\rm coul}.
}}
Here
$\r_{\rm coul} \equiv \r_{\rm qu} - [A_i^{\rm tr}, E_i^{\rm tr}]$ is the
color-charge density of the dynamical degrees of freedom, and
$M = M(A^{\rm tr}) = - D_i(A^{\rm tr})\p_i$ is the 3-dimensional Faddeev-Popov
operator.

	In this gauge, the Hamiltonian (without quarks) is given by
\eqn\ham{\eqalign{
  H & = (2g_0^2)^{-1} \int d^3x \ (E^2 + B^2)   \cr
	  & = (2g_0^2)^{-1} \int d^3x \ (E^{{\rm tr}2} + B^2)
+ (2g_0^2)^{-1} \int d^3x  \ d^3y \ \r_{\rm coul}(x)
\ {\cal V}(x,y; A^{\rm tr}) \ \r_{\rm coul}(y) \ .
  }}
Here
\eqn\pot{\eqalign{
{\cal V}(x,y; A^{\rm tr}) \equiv
    [M(A^{\rm tr})^{-1}(-\p^2)M(A^{\rm tr})^{-1}]_{\vec{x},\vec{y}}
}}
is a color-Coulomb potential-energy functional, depending on $A^{\rm tr}$, that
acts instantaneously and couples universally to color charge. The
continuum form of the Coulomb Hamiltonian, with proper attention to
operator ordering is given in \christlee\ and the lattice form in
\latcoulh.  For the {\it minimal} Coulomb gauge, this Hamiltonian is
supplemented by the boundary condition that the wave functionals
$\Psi(A^{\rm tr})$ are restricted to the Gribov region.  (This is
explained below and in the Appendix.)  It was proposed in
continuum theory in \gribov\ and in lattice theory in \critical,
that this restriction may be imposed by use
of an effective action or Hamiltonian
\eqn\effham{\eqalign{
H_{\rm eff} = H + (2g_0^2)^{-1} \int d^3x \
M^4 \ A_i^{\rm tr}(- \nabla^2)^{-1}A_i^{\rm tr} \ .
}}
As a result, the energy of a gluon of momentum $\vk$ gets modified to
$E^2(\vk) = \vk^2 + (\vk^2)^{-1}M^4$,
as one sees from the quadratic part of $H_{\rm eff}$.  One
obtains for the equal-time 3-dimensionally transverse would-be physical
gluon propagator the approximate expression
\eqn\gapprop{\eqalign{
D^{\rm tr}(\vk) = (2\pi)^{-1} \int dk_4
{ {1}\over {k_4^2 + E^2(\vk)} } = { {1}\over {2E(\vk)} }
}}
\eqn\approp{\eqalign{
D^{\rm tr}(\vk) =  { {|\vk|}\over { 2[(\vk^2)^2 + M^4]^{1/2} } }\ .
}}
This quantity is defined by
\eqn\trmom{\eqalign{
D_{ij}^{\rm tr}(\vk)
= (\d_{ij} - \hat{k}_i\hat{k}_j)D^{\rm tr}(\vk) \ ,
}}
where $D_{ij}^{\rm tr}(\vk)$ is the Fourier transform of
\eqn\trprop{\eqalign{
D_{ij}^{\rm tr}(\vx - \vy) =
\langle A_i^{\rm tr}(\vx, t) A_j^{\rm tr}(\vy, t)\rangle \ .
}}
In the absence of an estimate of corrections, one does not know how
accurate \approp\ may be.  However it was proven \vanish\
that the lattice
gluon propagator $D^{\rm tr}(\vk)$ at infinite spatial lattice volume
$L^3$ must indeed {\it vanish} at $\vk = 0$,
\eqn\suppress{\eqalign{
\lim_{\vk \to 0}D^{\rm tr}(\vk) = 0 \ ,
}}
although the rate of approach of
$D^{\rm tr}(0, L)$ to 0, as a function of $L$, was not established, nor
was it determined whether the renormalized gluon propagator also shares
this property.   The accuracy of \approp, and the crucial question of the
extrapolation to large
$L$ of $D^{\rm tr}(\vk, L)$ are addressed in the numerical study reported
here.

	We have also evaluated the 4-4
component of the gluon propagator
\eqn\ttprop{\eqalign{
D_{44}(\vx, t) \equiv \langle A_4(\vx, t)A_4(0, 0)\rangle
}}
at equal-time.  In the minimal Coulomb gauge $D_{44}(\vx, t)$
is given by~\rgcoul\
\eqn\ttinst{\eqalign{
D_{44}(\vx, t) = V(\vx)\d(t) + P(\vx, t) \ ,
}}
where $t = x_4$ is the Euclidean ``time" and $P(\vx, t)$ is a
non-instantaneous vacuum-polarization term.
This gives
\eqn\ttinstq{\eqalign{
\int_{-\e}^{+\e} dt \ D_{44}(\vx, t) = V(\vx) + o(\e),
}}
where $o(\e)$ vanishes with $\e$.
We call $V(\vx)$ the
color-Coulomb potential.  In
momentum space \ttinst\ reads
\eqn\ttmom{\eqalign{
D_{44}(\vk, k_4) = V(\vk) + P(\vk, k_4) \ ,
}}
where $\lim_{k_4 \to \infty}P(\vk, k_4) = 0$.  In dimension $d < 4$,
$V(\vx)$ coincides with
\eqn\expv{\eqalign{
V_0(\vec{x} - \vec{y}) \equiv \langle \ {\cal V}(x,y; A^{\rm tr}) \
\rangle. }}
However in $d = 4$ dimensions there is a mixing of $V(\vk)$ and
$P(\vk, k_4)$ associated with divergences, and $V(\vk)$ differs from
$V_0(\vk)$ by terms of the form $c_n g_0^{2n}/\vk^2$ in each order of
perturbation theory, as explained in detail in \rgcoul.

	Stated simply, the confinement of color is caused by the predominantly
long range of the color-Coulomb potential $V(\vx)$, corresponding to
an enhancement of $V(\vk)$ at low $|\vk|$.  Note however that $V(\vx)$ is
{\it not} the gauge-invariant energy eigenvalue of the quantum state of
infinitely massive separated quarks.  Nevertheless it is an important
quantity. It may be used as an order parameter for color confinement
\coul, and it is the starting point for calculations of the ground-state
wave-functional \cuzwsc, \Szcz\ and \Robertson.

	The rather surprising and counter-intuitive vanishing of
$D^{\rm tr}(\vk)$ at $\vk = 0$, and the enhancement
of $V(\vk)$ at $\vk = 0$  in
the minimal Coulomb gauge are both caused by the Gribov horizon.
This is a
boundary in the space of configurations $A_i^{\rm tr}(\vx)$  defined by the
condition that the Faddeev-Popov operator be positive, $M(A^{\rm tr}) \geq 0$.
The Gribov horizon represents the points where the lowest eigenvalue
$\l_0(A^{\rm tr})$ of $M(A^{\rm tr})$ first goes negative.  As shown in the
Appendix, all configurations that contribute to the Euclidean functional
integral in the minimal Coulomb gauge are constrained to lie within
the Gribov horizon.
Because of entropy considerations (see Refs.\ \gribov\ and
\critical), the Euclidean probability gets
concentrated near the horizon where the color-Coulomb interaction energy
$\cal{V}(A^{\rm tr})$, Eq.\ \pot, diverges.  This causes an enhancement
of the instantaneous color-Coulomb potential
$V(\vx)$.
 At the same time it places very severe  bounds on the magnitude of the low
momentum components of the gluon field.\foot{A similar confinement
scenario for the Landau gauge was proposed in \zbgr\ and \bgz.
It has also been verified numerically in the Landau gauge that typical
(thermalized and gauge-fixed) configurations lie very close to the
Gribov horizon \fmr.}

	The Coulomb gauge does not offer any particular advantage for
perturbative
calculations and renormalization.  However it is the finite limit of
renormalizable gauges~\recoul.  A valuable feature of this gauge is that the
time-time component of the gluon propagator
$D_{44}(\vk, k_4)$ is independent of both the cut-off $\L$ and the
renormalization
mass $\m$~\coul.  This holds separately for its instantaneous part
$V(\vk)$.  Thus the minimal Coulomb gauge allows us to introduce a running
coupling constant by
$g_{\rm coul}^2(|\vk|) = {\rm const} \  \vk^2 V(\vk)$.
The proportionality constant is determined
by the condition that in the
large-momentum or weak-coupling regime
$g_{\rm coul}^2(|\vk|)$
 satisfy the standard
renormalization-group equation,
\eqn\rge{\eqalign{
|\vk| { {\p g_{\rm coul} } \over {\p |\vk|} }
= \b_{\rm coul}(g_{\rm coul})
= - (b_0 g_{\rm coul}^3 + b_1 g_{\rm coul}^5 + \ldots) \
}}
where, for $SU(N)$ gauge theory without quarks,
$ b_0 = (4 \pi)^{-2} 11 N / 3 $,
$ b_1 = (4 \pi)^{-4} 34 N^2 / 3$.
The proportionality constant is calculated in \rgcoul, with the result
that for $SU(N)$ gauge theory without quarks
\eqn\propco{\eqalign{
\vk^2V(\vk) = {  {12} \over {11} } \
g_{\rm coul}^2(|\vk|/\L_{\rm coul}) \ ,
}}
and more generally, with $N_f$ quark flavors
\eqn\propcoq{\eqalign{
\vk^2V(\vk) = {  {12 N} \over {11N - 2N_f} } \
g_{\rm coul}^2(|\vk|/\L_{\rm coul}) \ .
}}
Here $\L_{\rm coul} \propto \L_{\rm QCD}$ is a finite QCD mass scale,
characteristic of the Coulomb gauge, such that asymptotically in the
weak-coupling regime
 \eqn\rg{\eqalign{
\vk^2 = \L_{\rm coul}^2 \ \exp[(b_0 g_{\rm coul}^2)^{-1}]
 \ (b_0 g_{\rm coul}^2)^r
   \ , }}
where $r \equiv b_1/b_0^2$.  The ratio
$\L_{\rm coul}/\L_{\rm QCD}$ may be obtained from a 2-loop
calculation \rgcoul.  We conclude that in the minimal Coulomb gauge the
running coupling constant of QCD may be obtained from a numerical
determination of the equal-time 2-point function $D_{44}$,
whereas in other gauges
it must be obtained from a 3-point function.  The running coupling
constant $g_{\rm coul}$ that we have introduced is the QCD analog of the
invariant charge in QED that is defined in terms of the transverse part of
the photon propagator in a Lorentz-covariant gauge.

\newsec{Results}

We evaluate the space-space gluon propagator
\eqn\dkdef{\eqalign{
D^{\rm tr}(\vec{0})& = {1 \over 9 V} \sum_{t = 1}^{L}
 \sum_{\mu = 1}^{3}
  \sum_{b = 1}^3 \, D_{\mu \mu}^{b b}(\vec{0},t) \cr
D^{\rm tr}(\vk) & = {1 \over 6 V} \sum_{t = 1}^{L}
   \sum_{\mu = 1}^{3}
  \sum_{b = 1}^3 \, D_{\mu \mu}^{b b}(\vk,t) \
,  }}
and the time-time gluon propagator
\eqn\dkdeftt{\eqalign{
D_{44}(\vk) = {1 \over 3 V} \sum_{t = 1}^{L}
   \sum_{b = 1}^3 \, D_{44}^{b b}(\vk,t) \
, }}
where
\eqn\dmu{\eqalign{
 D_{\mu \mu}^{b b}(\vk,t) = \langle\,
 \left\{\,\left[\,\sum_{\vx}\,A_{\vx t \mu}^{b}\,
 \cos{( \vec{\t} \cdot \vx )}\,\right]^{2}
  + \left[\,\sum_{\vx}\,A_{\vx t \mu}^{b}\,
 \sin{( \vec{\t} \cdot \vx )}\,\right]^{2}
 \, \right\} \,\rangle \ ,}}
and the lattice gluon field $A_{\vx t \mu}^{b}$ is defined in
Eq.\ (A.5).
Here, as usual, we have defined $k_i \equiv 2 \sin(\theta_i/2)$, for
$ -\pi \leq \theta_i = 2 \pi n_i/L \leq \pi$, and integer $n_i$.
An average over the $L$ time slices is included.
In our simulations we consider only 3-momenta aligned along major axes
$\theta_i = (0, 0, 2\pi n/L)$.  Notice that $D^{\rm tr}(\vec{0})$
is not given by $D^{\rm tr}(\vk)$
at $\vk = \vec{0}$. The difference is due to
the Coulomb gauge condition --- the continuum-like condition,
Eq.\ (A.4) --- which in momentum space reads
\eqn\pAfourier{\eqalign{
\sum_{i = 1}^{3}\, k_{i} \, {\widetilde A}_{\vk t i}
 \, =\, 0
 \ ,}}
where ${\widetilde A}_{\vk t i}$ is the three-dimensional
Fourier transform of the gauge field $ A_{\vx t i} $.
If $\vk \neq (0,\, 0,\, 0)$ only two
of the three Lorentz components of
${\widetilde A}_{\vk t i}$ --- and therefore of $A_{\vx t i}$ ---
are independent.
This explains the factor $6$ (instead of $9$) in
the definition of $D^{\rm tr}(\vk)$.

Fig.\ 1 shows $D^{\rm tr}(\vk)$ and $D_{44}(\vk)$ as a function of
$\vk^2$ on the same logarithmic plot for the lattice sides $L = 28$ and $30$.
The qualitative behavior is quite different in the two cases.  Whereas
$D_{44}(\vk)$ grows strongly at low $\vk$, by contrast $D^{\rm tr}(\vk)$ turns
over and decreases at low $\vk$.

(a) {\bf Analysis of $D^{\rm tr}$}.   Figs.\ 2, 3 and 4 show our data
for $D^{\rm tr}(\vk)$ for the nine lattice volumes considered. To
parametrize these data, we were guided by the Nielsen identities
\nielsen.  They tell us that the poles of the gluon propagator
$D^{\rm tr}(\vk, k_4)$ are independent of the gauge
parameters.\foot{Strictly
speaking, the Nielsen identities have been established only for
Faddeev-Popov type gauge fixing.  This does not include the minimal
Coulomb gauge because the
gauge fixing is done by a minimization procedure that is not
describable by a local 4-dimensional action.
However the minimal Coulomb gauge may be obtained as a
limiting case of a local 5-dimensional quantum field theory that
describes stochastic quantization with stochastic gauge fixing, as has
been discussed recently in \zbgr\ and \bgz.  The Nielsen identities may be
extended to the 5-dimensional formulation.}    In particular, for the
class of gauges defined by
$\lambda\, \p_4A_4 + \p_iA_i =
0$, which interpolate between the Landau gauge, $\l = 1$, and the Coulomb
gauge,
$\l = 0$, the poles are independent of $\l$.  In the Landau gauge, the
poles occur at $k^2 = \vk^2 + k_4^2 = - m^2$, by Lorentz (Euclidean)
invariance, and thus also
in all these gauges, by virtue of the Nielsen identities.  We have made a
2-pole fit
with poles at $m_1^2$ and
$m_2^2$, which may be either a pair of real numbers or a complex conjugate
pair.\foot{According to the general
principles of quantum field theory, the propagator of physical particles
should have poles only at real
positive $m^2$.  However in the confined phase the gluon propagator may
have singularities that correspond to unphysical excitations. }
According to
\vanish, the propagator in the minimal Coulomb gauge vanishes at $\vk = 0$.
We
use this condition to fix the residues to within an over-all normalization
\eqn\pole{\eqalign{ D^{\rm tr}(\vk, k_4) & =
C \ (\vk^2)^{\a-1} \  \Big({{k_4^2 + m_1^2} \over {\vk^2 + k_4^2 + m_1^2}}
- {{k_4^2 + m_2^2} \over {\vk^2 + k_4^2 + m_2^2}} \Big)   \cr
& =
-\, C \ (\vk^2)^{\a} \  \Big({1 \over {\vk^2 + k_4^2 + m_1^2}}
- {1 \over {\vk^2 + k_4^2 + m_2^2}} \Big)  \ , }}
where $\a$ is a fitting parameter.  This formula does not reproduce the
correct asymptotic behavior at large momenta $\sim (\vk^2 + k_4^2)^{-1}$
times logarithmic corrections.  However, this
is not a problem because our largest value of $|\vk|$ is less than 2
GeV, and we are probably far from the ultraviolet regime.\foot{In
the Landau gauge \gluonasym\ the gluon propagator reaches three-loop
asymptotic scaling at momenta of about $5$--$6$ GeV.}
We wish to emphasize that our pole parameters --- to the extent
that they are a valid fit --- are gauge-independent quantities that
characterize the gluon.

	The equal-time part of this propagator is
given by
$D^{\rm tr}(\vk) = (2\pi)^{-1}\int d \theta_4 \ D^{\rm tr}(\vk, k_4)$,
where $k_4 = 2 \sin(\theta_4/2)$.  This gives
 \eqn\eqtime{\eqalign{ D^{\rm tr}(\vk) & =
- \, C^{\prime} \ (\vk^2)^{\a} \  (A_1^{-1/2} - A_2^{-1/2})   \cr
& =  C^{\prime \prime} \ (\vk^2)^{\a} \  {{4 + h_1 + h_2} \over
{A_1^{1/2}A_2^{1/2}(A_1^{1/2} +A_2^{1/2})}}
   \ , }}
where $h_i \equiv \vk^2 + m_i^2$ and $A_i \equiv 4 h_i + h_i^2$ for $i =
1,2$.  If
we take a pair of complex conjugate poles, $m_1^2 = x + iy$ and $m_2^2 = x
- iy$,
we obtain
 \eqn\cmplx{\eqalign{ D^{\rm tr}(\vk) =  C^{\prime \prime \prime} \
(\vk^2)^{\a} \
{{v} \over {(u^2 + 4y^2v^2)^{1/2}\, [(u^2+4y^2v^2)^{1/2} +u]^{1/2}}}
   \ , }}
where
$u \equiv 4(\vk^2 + x) + (\vk^2 + x)^2 - y^2$ and
$v \equiv (2 + \vk^2 +x)$.  The case of a pair of real poles is obtained
from this formula by taking a negative value of $y^2$.  For the case of a
lattice of finite volume $V = L^4$, we modify this formula to
 \eqn\mod{\eqalign{ D^{\rm tr}(\vk) =  z \ [(\vk^2)^{\a} + r] \
{{v} \over {(u^2 + 4y^2v^2)^{1/2} \, [(u^2+4y^2v^2)^{1/2} +u]^{1/2}}}
   \ . }}
The factor $v[(u^2+4y^2v^2)^{1/2} +u]^{-1/2} =
(A_1 - A_2)[8^{1/2}iy(A_1^{1/2} +A_2^{1/2})]^{-1}$ is slowly varying over
the relevant range of $\vk$ and parameter values, so a fit to \mod\ is
a test of the simpler formula
\eqn\simp{\eqalign{ D^{\rm tr}(\vk) =  z' \ [(\vk^2)^{\a} + r] \
{{1} \over {(u^2 + 4y^2v^2)^{1/2}}}
   \ . }}
Stated differently,
\mod\ and \simp\ have the same singularities that are nearest to
the origin.
In the continuum limit we have $u \to 4(\vk^2 + x)$ and $v \to 2$, and
\simp\ is a lattice discretization of Gribov's approximate formula
\approp\ provided that the fitting parameters have the values
$r = 0$, $\a = 0.5$, $x = 0$, with the identification $y^2 = M^4$.  It
is intended report on the fit to \simp\ elsewhere, but preliminary
indications are that it is comparable in quality to the fit to \mod.

	For each lattice side $L$ we have made a fit of the parameters $z(L),\,
r(L),\, \a(L),\, x(L)$ and $y^2(L)$.
By using Table 3 of \stringtens$\,\!$ and by
setting the physical string tension equal to
$\sqrt{\sigma} = 0.44$ GeV we obtain that, for $\b = 2.2$, the
inverse lattice spacing is $a^{-1} = 0.938$ GeV. This gives
$a = 0.21$ fm, so that the largest lattice volume
considered here, i.e.\ $V = 30^4$, corresponds to $(6.3$ fm$)^4$, the
smallest non-zero momentum that can be
considered for that lattice is equal to $0.196$ GeV, while
the maximum momentum value (for each lattice side $L$) is 1.876 GeV.
The results\foot{The fits have been done using {\tt gnuplot}; the
errors represent $68.3\%$ confidence interval. Similar
results have been obtained using a conjugate-gradient
method with errors estimated by a jack-knife method.
Let us notice that we did not consider the correlations
between different momenta when fitting the data. However,
at least in the Landau gauge, the covariance matrix for the
gluon propagator in momentum space is essentially diagonal
\bps, and the value of the ``naive'' $\chi^2/$d.o.f. is
usually compatible with the value obtained using the full
covariance matrix \gluonadelaide.}
are exhibited in Table 1, and the curves
are plotted in Figs.\ 2, 3 and 4.
There is no a priori reason why a 2-pole fit should be
accurate over the whole range of momenta considered.
However the fit is excellent for all momenta $\vk^2$ and
for each $L$. We have also checked that results in
agreement with those reported in Table 1 are obtained
if one considers only the data corresponding to $\vk^2 < 2$.

We have extrapolated to infinite $L$ the
fitting parameters $z(L),\, \alpha(L),\, x(L),\, y^2(L)$ and the product
$(rz)(L)$;
in all cases we tried three different types of fitting functions,
namely $a + b/L^c$, $\exp(a)/L^b$ and $a + \log(1 + b/L)$, and chosen
the fit with smallest $\chi^2/$d.o.f. Results\foot{These fits
have also been done using {\tt gnuplot}. For the fitting function
$\exp(a)/L^b$ the results have been checked with the exact
minimizing formula (notice that this fitting function is
linear in the coefficients $a$ and $b$
after taking the logarithm).} are
shown in Table 2 (first row) and plotted in Figs.\ 5--9.

	The reader will have noticed that the product
$(rz)(L)$ extrapolates to
0 (see Fig.\ 9). This corresponds to a {\it vanishing} of
$D^{\rm tr}(0)$ at infinite lattice volume, in accordance with \vanish.
To check on this important point we have also fitted $D^{\rm tr}(0, L)$
using the three fitting functions considered above.
A good fit is provided by $D^{\rm tr}(0, L) = \exp(a) / L^b$,
with $a = 2.43(4)$, $b = 0.50(1)$,
$\chi^2/$d.o.f.\ $= 0.69$ and goodness-of-fit $Q = 65.6 \%$
(see Fig.\ 10).\foot{When using the fitting function $a + b/L^c$ we
obtain, both for $(rz)(L)$ and $D^{\rm tr}(0, L)$, a value of
$a$ that is zero within errors but a worse $\chi^2/$d.o.f.}
An indication of how reliable these fits are is the
comparison of the power $b$ in the fit of the product $(rz)(L)$ and of
$D^{\rm tr}(0,L)$ which are $b = 1.1(1)$ and $b = 0.50(1)$ respectively.
We have also made a similar fit for $r(L)$ (not plotted)
and obtained $b = 1.8(1)$.

	We next consider the $\vk$ dependence of $D^{\rm tr}(\vk)$ at low
$\vk$.  For the power dependence parametrized by $(\vk^2)^{\a}$, observe
that $\a(L)$ extrapolates to $\a(\infty) = 0.48(5)$, see Fig.\ 6.
This agrees with Gribov's approximate formula \approp, which
gives $\a = 0.5$.  Moreover this value is consistent with the other rows
of Table 2, by the method described below, so this result appears quite
stable.  Particularly striking is that, with $x = 0$ imposed, one
obtains  $\a = 0.49(1)$ and $\a = 0.51(1)$ respectively from the fourth
and fifth rows of Table~2 (explained below).

	Another striking feature of the fit is that $y^2(L)$ is positive
for all
9 values of $L$ (see Table 1),
corresponding to a pair of {\it complex} conjugate poles rather than a pair
of real
poles. Moreover, in all cases $x(L)$ is quite small compared to $y(L)$.  For
example $x(30) = 0.06(6)$ and $y(30) = 0.88(5)$. The extrapolation to
infinite $L$ also strongly indicates a positive value $y^2(\infty) > 0$
and a small, possibly zero, value for $x(\infty)$.
Thus our data are compatible with and perhaps
suggestive of poles
at purely imaginary $m^2 = 0 \, \pm \, i y$, in agreement with Gribov's
formula \approp\ with $y^2 = M^4$.

	Finally, we have fit the data for $L = 28$ and $L = 30$ using
Eq.\ \mod\ with $r = 0$ and with a low-momentum cut $\vk^{2}_{\rm min}$,
namely considering only a range of momenta in which
finite-size effects are negligible. Results are reported in the second
and third rows of Table 2 for two different values of
$\vk^{2}_{\rm min}$. Similar results are also obtained when the fixed
value
$x = 0$ is imposed (see the last two rows of Table 2 and Fig.\ 11).
The values for $\alpha$ and $y^2$ obtained in this way
are in good agreement with the values obtained by extrapolating
$\alpha(L)$ and $y^2(L)$ to infinite $L$ (compare the first row of
Table 2 with the other four rows of the same table).

	(b) {\bf Analysis of $D_{44}$}.
Eq. \ttinstq\ shows that for lattice quantities we may make the
identification
$V(\vx) = D_{44}(\vx)$, where $D_{44}(\vx)$ is the equal-time propagator,
and similarly for their 3-momentum transforms, $V(\vk) = D_{44}(\vk)$.
This allows us to use \dkdeftt\ and \propco\ to  also define the
lattice quantity $g_{\rm coul}^2(\vk)$ by
\eqn\lrcc{\eqalign{
{ {12} \over {11} } g_{\rm coul}^2(\vk)
\equiv \vk^2 V(\vk) \equiv \vk^2 D_{44}(\vk) \ . }}

	Fig.\ 12 shows our data for the running coupling constant
${ {12} \over {11} } g_{\rm coul}^2(\vk)$.  In the continuum limit, its
behavior
at large momentum is governed by the perturbative renormalization group
\rge\ and \rg.
For the fitting formula we modify \rg\ to
\eqn\mrg{\eqalign{ \vk^2 = \L_{\rm coul}^2
 \ \exp[(b g_{\rm coul}^2)^{-1}]
 \ [(b g_{\rm coul}^2)^{-r}
+ z  (b g_{\rm coul}^2)^{\a}]^{-1} \ , }}
which implicitly defines $g_{\rm coul}^2(\vk^2)$.  Here $r = 102/121$, and
$ \L_{\rm coul}^2, b, z, \a$
are fitting parameters whose significance we now explain.  Naturally the
parameter
$\L_{\rm coul}^2$ sets the mass scale.  For small $g_{\rm coul}^2$,
which corresponds to large $\vk^2$, this formula is dominated by the first term
in the denominator, whereas for large $g_{\rm coul}^2$, which corresponds to
small $\vk^2$, it is dominated by the second term in the denominator.
For small $g_{\rm coul}^2$ the formula approaches \rg, provided that
$b = b_0 = { {11} \over {24 \pi^2} } \approx 0.046$.
However we are quite far from the continuum limit for
$D_{44}(\vk)$, as indicated by the small value
$\langle (1/2) {\rm tr}(U_4)\rangle = 0.221(6)$ (for lattice volume
$14^4$),\foot{Before minimizing $F_{\rm ver}$ we find
$\langle (1/2) {\rm tr}(U_4)\rangle = -0.0005(57)$.} and we
expect significant $\b$-dependence in the extrapolation to the continuum
limit.
Moreover, for fixed $\beta$, it has been found that
different lattice discretizations of the gluon field
lead to identical gluon propagators to within numerical accuracy, apart
from the overall normalization \discreti.
We allow for this by taking the overall
normalization of $g_{\rm coul}^2$ to be a fitting parameter.  This requires
putting
an arbitrary normalization coefficient everywhere in front of
$g_{\rm coul}^2$ in
\mrg, which is equivalent to replacing the fixed number
$b_0$ by  the fitting parameter $b$. Of course, an extrapolation
in $\b$ to
the continuum limit should give $b = b_0$.

	For large $g_{\rm coul}^2$ and small $\vk^2$, this formula approaches
\eqn\ir{\eqalign{
	b g_{\rm coul}^2
= \Big( { { \L_{\rm coul}^2 }  \over { z \vk^2 } }\Big)^{1/\a} \ .
}}
Thus the parameter $z$ sets the overall normalization in the strong-coupling or
infrared regime, and $\a$ governs the strength of the singularity of
$g_{\rm coul}^2$ in the infrared limit.  If the color-Coulomb potential
$V(\vk)$ is governed by a string tension at large distances then
$g_{\rm coul}^2(\vk) \sim {\rm const}/\vk^2$, which corresponds to $\a = 1$.

	On a finite periodic lattice with finite lattice spacing
and with $g_{\rm coul}^2(\vk)$ defined by
${ {12} \over {11} } g_{\rm coul}^2(\vk) \equiv \vk^2 V(\vk)$,
necessarily $g_{\rm coul}^2(\vk)$ vanishes at $\vk = 0$.
Actually $g_{\rm coul}^2(\vk)$ gets a maximum value at
$\vk^{2} \approx 0.2$ and goes to zero in the infrared limit
(see Fig.\ 12). This unphysical behavior is a lattice artifact that also
appears in
studies of the running coupling constant using the three-gluon vertex
\gluonalf. In order to fit our data we have made a
low-momentum cut at $\vk^{2}_{\rm min} = 0.5$.
The values of the parameters which we obtain\foot{The fit has been
done using a conjugate gradient method with a numerical inversion of
Eq.\ \mrg. Errors are estimated using a jack-knife method.}
are $\L_{\rm coul}^2 = 1.0(2)$, $z = 1.2(1)$, $b' = 0.18(2)$,
$b = 0.20(2)$, $\a = 1.9(3)$, with
$\chi^2/$d.o.f.\ $= 0.44$ and goodness-of-fit $Q = 98.3 \%$
using the data for $L = 28$ and $L = 30$,
where $b' \equiv { {11} \over {12} } b$. We have checked that
similar results are obtained with $\vk^{2}_{\rm min} = 0.3$ and
$\vk^{2}_{\rm min} = 1.0$ (see Table 3) but the resulting
$\chi^2/$d.o.f.\ is smallest for $\vk^{2}_{\rm min} = 0.5$.

The value of $\a \sim 2$ corresponds to
$g_{\rm coul}^2 \sim {\rm const}/|\vk|$, and
$V(\vk) \sim {\rm const}/|\vk|^3$ at low momentum.
The volume dependence of the data, and therefore of our fit,
is quite weak, but one notices in Table 3 that the value of $\a$
decreases as the low momentum cut-off $k_{\rm min}^2$
increases. This corresponds to an increase in the
strength of the singularity of $g_{\rm coul}^2(\vk)$ at $\vk = 0$
as finite-volume effects are reduced.
However, as explained above, we must make an extrapolation in $\b$ in
order to arrive at any precise conclusion about the strength of the
singularity in the continuum limit. Nevertheless, our data
at finite $\b$ and $L$ clearly
indicate a color-Coulomb potential that is more singular than
$V(\vk) \sim {\rm const}/|\vk|^2$ at low $\vk$.

\newsec{Conclusions}

We have used simple formulas to fit the data for the equal-time gluon
correlators
$D^{\rm tr}(\vec{k})$  and $D_{44}(\vec{k})$  in $SU(2)$ lattice gauge theory
in the
minimal lattice Coulomb gauge at $\beta = 2.2$.
Our fits have the following features:

1.\ The equal-time would-be physical gluon propagator $D^{\rm tr}(\vk, L)$
at  {\it zero} momentum $\vk$ extrapolates to 0 in the limit of infinite
lattice volume, i.e.\
$D^{\rm tr}(0, \infty) = 0$.  The rate of approach for lattice volume
$V = L^4$ was fit by $D^{\rm tr}(0, L) = C L^{-b}$, where $b =
0.50(1)$. The {\it vanishing} of the gluon propagator
$D^{\rm tr}(\vk)$ at $\vk = 0$ is highly counter-intuitive, and the only
explanation
for it is the suppression of the low momentum components of the gluon field
that is a particular feature of the Gribov horizon, that constitutes
the boundary of configuration space.\foot{A similar
result, i.e.\ a transverse gluon propagator that
at {\it zero} momentum $\vk$ extrapolates to 0 in the limit of infinite
lattice volume $V$, has been recently obtained for pure $SU(2)$
lattice gauge theory in the three-dimensional case
and in the magnetic sector at finite temperature
\gluonhot.}

2.\  Asymptotically at low momentum our fitting formula behaves like
$D^{\rm tr}(\vec{k}) \propto (\vk^2)^\a$, with the value extrapolated to
infinite lattice volume $\a = 0.48(5)$ (see Table 2).  With the value
$x = 0$ imposed and a low-momentum cut-off one obtains (from $L = 28$ and
$L = 30$) $\a = 0.49(1)$ or $\a = 0.51(1)$ (see Table 2).  This is in
striking numerical agreement with Gribov's formula
$D^{\rm tr}(\vk) = (|\vk|/2)[(\vk^2)^2 + M^4]^{-1/2}$, which gives
$\a = 0.5$.

3.\  We have obtained an excellent 2-pole fit for $D^{\rm tr}(\vec{k})$.
Our fit
indicates that the poles occur at {\it complex} $ m^2 = x \, \pm \, i y$.  The
real part $x$
is quite small and compatible with 0.  Remarkably, a pole in $k^2$ at
purely imaginary $m^2 = 0 \, \pm \, i y$ agrees
with the Gribov propagator
$D^{\rm tr}(\vk) = (|\vk|/2)[(\vk^2)^2 + M^4]^{-1/2}$,
with $y^2 = M^4$.  Note that only a
purely imaginary pair of poles gives a correction to the free equal-time
propagator
$D^{\rm tr}(\vk)
\approx |\vk|^{-1}(1 - \demi { {M^4} \over {(\vk^2)^2} })$ of relative
order $(\vk^2)^2$ with coefficient of dimension (mass)$^4$.  It may
not be a coincidence that this is the dimension of the gluon condensate
$\langle F^2 \rangle$, which is the lowest dimensional condensate in
QCD.  Because of the gauge
invariance of the location of the poles, by virtue of the Nielsen
identities \nielsen, and because of the theoretical suggestiveness of our
result, we are encouraged to report the values
$m^2 = 0 \, \pm \, i y$, for
$y = 0.671(7)$ in lattice units, or $y = 0.590(7)$~GeV$^2$,
$M = y^{1/2} = 0.768(4)$~GeV for the location of the gluon poles in $k^2$.

4.\ The Coulomb gauge offers a definition of the running coupling
constant,
${ {12} \over {11} } g_{\rm coul}^2(\vk) = \vk^2 D_{44}(\vk)$,
which has advantages for numerical
determination.  It is less subject to fluctuation than the determination of
$g^2$ from
the 3-point function in the Landau gauge \gluonalf, but the
extrapolation in $\b$ remains to be done.

5.\ Our data for $D_{44}(\vec{k})$
require a cut at low momentum  to eliminate lattice artifacts.  After
this cut, the running coupling constant defined by
${ {12} \over {11} } g_{\rm coul}^2(\vec{k}/\Lambda_{QCD})
= \vec{k}^2 D_{44}(\vec{k})$
extrapolates to low momentum in accordance with infrared slavery, namely
the running coupling constant $g_{\rm coul}^2(\vec{k}/\Lambda_{QCD})$
diverges in the zero-momentum limit.

6.\   The observed strong {\it enhancement} of the instantaneous
color-Coulomb potential $V(\vk)$ and the strong {\it suppression} of the
equal-time would-be physical gluon propagator $D^{\rm tr}(\vk)$ both at
low $\vk$, strongly
support the confinement scenario of Gribov \gribov, \coul.  In addition
to this qualitative agreement, we note excellent numerical
agreement of our fit to Gribov's formula
$D^{\rm tr}(\vk) = (|\vk|/2)[(\vk^2)^2 + M^4]^{-1/2}$, reported in 1, 2,
and 3 above.  If this excellent fit is maintained at larger $\b$ values,
then it appears that we have obtained a quantitative understanding of
$D^{\rm tr}(\vk)$.

\vskip 3mm

  \centerline{\bf Acknowledgments}

The research of Attilio Cucchieri was partially supported by
the TMR network Finite Temperature Phase Transitions in
Particle Physics, EU contract no.: ERBFMRX-CT97-0122.
The research of Daniel Zwanziger was partially supported by the National
Science Foundation under grant PHY-9900769.

\appendix A{Numerical gauge fixing}

	The minimal lattice Coulomb gauge is defined by two gauge-fixing
steps.  In
the first step the {\it spatial } link variables
$U_{\vx t i} \in SU(N)$, for $i = 1, 2, 3$, are made as close to unity as
possible by
minimizing the ``horizontal" minimizing function
\eqn\hfunct{\eqalign{ F_{{\rm hor},U}(g) \equiv
 \sum_{\vx,t,i} {\rm Re \ Tr}(1 - {^g}U_{\vx ti}) \ , }}
with respect to gauge transformations $g_{\vx t}$.   The sum extends over
all horizontal or space-like links, and the minimization is done
independently on each time-slice $t = x_4$.  (The gauge transform
${^g}U_{x\m}$ of the link variable $U_{x\m}$ is defined by ${^g}U_{xy} \equiv
g_x^{-1}\, U_{xy}\, g_y$, where $y \equiv x + \hat{\m}$ and
$\hat{\m}$ is the unit vector in the $\m$ direction.)  The stochastic
over-relaxation
algorithm does not necessarily yield the absolute minimum of the minimizing
function but leads in general to one of several local minima.  Different
minima correspond to different Gribov copies.
For the lattice volumes $16^4$ and $20^4$ we have checked that
the dependence of the gluon propagators $D^{\rm tr}(\vk)$ and $D_{44}(\vk)$
on which Gribov copy one ends up is of the order of magnitude
of the numerical accuracy, in agreement with Ref.\ \gnoise.

After this step the lattice gluon field is
3-dimensionally transverse [see Eq.\ (A.4) below].
But the gauge fixing is as yet incomplete
because it leaves a $t$-dependent but $\vx$-independent gauge
transformation $g_t$
arbitrary.   This arbitrariness is fixed in the second step in which the
{\it time-like}
link variables $U_{\vx t 4}$ are made as close to unity as possible by
minimizing the
``vertical" minimizing function
\eqn\vfunct{\eqalign{ F_{{\rm ver},U}(g) \equiv
 \sum_{\vx,t} {\rm Re \ Tr}(1 - {^g}U_{\vx t4}) \ , }}
with respect to $\vx$-independent gauge transformations
$g_t$.  The sum extends over all vertical or time-like links.  This gauge
fixing of
the vertical links does not alter the spatial correlator $D^{\rm tr}(\vk)$
nor the
color-Coulomb potential $\cal{V}(A^{\rm tr})$.  However it reduces the
non-instantaneous part of $D_{44}(\vk)$,
so that it is suppressed compared to the
instantaneous part $V(\vk)$ and vanishes with the lattice spacing in the
continuum limit.

This gauge fixing, in which first $F_{\rm hor}$ is minimized and then
$F_{\rm ver}$ is minimized is equivalent to the limit of the interpolating
gauge in which the single function, depending on a real positive parameter
$\lambda$,
\eqn\lambdagauge{\eqalign{
        F = F_{\rm hor} + \lambda F_{\rm ver}
}}
is minimized, and the limit $\lambda \to 0$ is taken.

	With this gauge fixing, the link variables should approach unity in
the continuum
limit in the sense that
$\lim_{\b \to \infty}(1/2){\rm Tr}(U_{x \m}) = 1$.  In our study at
$\b = 2.2$ we obtain
$\langle(1/2){\rm Tr}(U_{x i})\rangle = 0.86249(3)$ for the space-like links,
whereas for time-like links
$\langle(1/2){\rm Tr}(U_{x 4})\rangle = 0.221(6)$ on lattice volume $14^4$.
(The
dependence on the volume is not strong.)  Thus we are quite far from the
continuum
limit for quantities such as
$V(\vk)$ that depend on the vertical links, and they may exhibit significant
$\b$-dependence in the extrapolation to the continuum limit.  This will be
reported subsequently.

	The gauge fixing just described produces a configuration $U$ which
is a local minimum of $F_{{\rm hor},U}(g)$ and $F_{{\rm ver},U}(g)$
at $g = 1$.  At a local minimum the minimizing functions are 1)
stationary under infinitesimal variations $\d F = 0$ and 2) the
matrix of second variations of the minimizing function is
non-negative, $\d^2 F \geq 0$.  We now comment on implications of
these properties for gauge-fixed configurations $U$.  At a local
minimum, the horizontal minimizing function
$F_{{\rm hor},U}(g)$ is stationary with respect to infinitesimal variations
$g_x \to
g_x(1 +
\om_x)$.  Here
$\om_x = t^a \om_x^a$ is an element of the Lie algebra of the $SU(N)$
group, with
anti-hermitian basis $t^a$ satisfying $[t^a, t^b] = f^{abc}t^c$ and
${\rm Tr}(t^a t^b) = - \demi \d^{ab}$.  The corresponding variation of
$F_{\rm hor}$ is given by
\eqn\stat{\eqalign{ \d F_{{\rm hor},U}(g) =
 - \demi \sum_{\vx,i} {\rm Tr}[ (\om_{\vx + \hi,t} - \om_{\vx t})
({^g}U_{\vx t i} - {^g}U_{\vx t i}^{\dag} ) ] \ .
}}
For a configuration $U$ which is a local minimum (at $g_x = 1$), this quantity
must vanish for all
$\om_{\vx t}$, which gives
\eqn\trans{\eqalign{
  \sum_i (A_{\vx,t,i} - A_{\vx - \hat{i},t,i}) = 0 \ .
}}
Here $A_{\vx,t,i}$ is the lattice gluon field defined by
\eqn\deflgf{\eqalign{
A_{x\m}^a \equiv - {\rm Tr}[t^a(U_{x\m} - U_{x\m}^{\dag})],
}}
which is a lattice analog
of the continuum connection $A_\m^a(x)$.  Equation \trans\
 is the lattice
transversality condition for spatial directions, which is the defining
condition for the lattice Coulomb gauge.

	Since the gauge-fixed configuration $U$ is a local minimum of
$F_{{\rm hor},U}(g)$ (at $g_x = 1$), its second variation is non-negative
\eqn\pos{\eqalign{
\d^2 F_{{\rm hor},U}(g) = (\om, M(U) \om) \geq 0 \
 \ \ \ {\rm for \ all \ }\om \ .
}}
Here $M(U)$ is the lattice Faddeev-Popov matrix defined on a given
time-slice $t$ by
\eqn\FP{\eqalign{
(\om, M(U) \om) \equiv
 - \demi \sum_{\vx,i} {\rm Tr}[ (\om_{\vx + \hi} - \om_{\vx})
(U_{\vx i}\,\om_{\vx + \hi} - \om_{\vx}\,U_{\vx i}
+ \om_{\vx + \hi}\, U_{\vx i}^{\dag} - U_{\vx i}^{\dag}\, \om_{\vx}) ] \ ,
}}
and we have suppressed the index $t$ which is common to all
variables.  The positivity of $M(U)$
is a condition on configurations $U$ which, together with the
transversality
condition \trans, defines the lattice Gribov region, whose boundary is the
Gribov horizon.

	[That this is a highly restrictive condition is suggested by the
following consideration.  The
Faddeev-Popov matrix $M(U)$ for $SU(N)$ gauge theory
is a symmetric matrix of dimension
$\,V(N^2-1) $, where $V$ is the (large) number of sites of the lattice, so it
has $\,V(N^2-1)\,$ eigenvalues.  Configuration space is divided into
$\,V(N^2-1) + 1\,$ different regions $R_n$ according to the number $n = 0,
\ldots,
V(N^2-1) \,$ of positive eigenvalues of $M(U)$.  Of these, the Gribov region
consists of the single region $R_{V(N^2-1)}$ that includes $U_{\vx i} = 1$.
For
the $SU(2)$ group at least, all regions are populated.  To see this, observe
that for
$U_{x\m} = 1$, we have
$M(1) = - \D$, whereas for
$U_{x\m} = -1$, we have
$M(-1) = \D$, where $\D$ is the lattice Laplacian. In these 2 cases,
depending on the
sign, the configuration
$U = \pm 1$ is in region
$R_{V(N^2-1)}$ or $R_0$.  By continuity therefore all
$\,V(N^2-1) + 1\,$ different regions are populated.  Similar considerations
apply to
$F_{{\rm ver},U}(g)$.]


\footatend\vfill\supereject\immediate\closeout\rfile\writestoppt
\baselineskip=14pt\centerline{{\bf References}}\bigskip{\frenchspacing%
\parindent=20pt\escapechar=` \input refs.tmp\vfill\eject}\nonfrenchspacing


\vfill\eject\immediate\closeout\tfile{\parindent40pt
\baselineskip14pt\centerline{{\bf Table Captions}}\nobreak\medskip
\escapechar=` \input tabs.tmp\vfill\eject}

\topinsert
$$\vbox{\halign{
\hfil#\hfil& \quad\hfil#\hfil& \quad\hfil#\hfil&
 \quad\hfil#\hfil& \quad\hfil#\hfil& \quad\hfil#\hfil&
 \quad\hfil#\hfil& \quad\hfil#\hfil\cr
\noalign{\hrule}
\ \cr
$L$ & $z$ & $r$ & $\alpha$ & $x$ & $y^2$ & $\chi^2/$d.o.f.\ & $Q$ \cr
\ \cr
\noalign{\hrule}
\ \cr
$14$ & $3.93(33)$ & $3.86(35)$ & $0.76(5)$ & $0.36(6)$  & $1.07(5)$  & $0.42$ &
$73.9\%$ \cr
$16$ & $4.79(42)$ & $2.72(28)$ & $0.70(5)$ & $0.27(8)$  & $1.08(7)$  & $1.46$ &
$21.1\%$ \cr
$18$ & $5.91(53)$ & $2.27(33)$ & $0.57(7)$ & $0.38(13)$ & $1.11(13)$ & $3.45$ &
$0.4\%$ \cr
$20$ & $5.78(19)$ & $1.80(8)$  & $0.66(2)$ & $0.19(4)$  & $0.99(4)$  & $0.45$ &
$84.5\%$ \cr
$22$ & $5.81(23)$ & $1.57(8)$  & $0.67(3)$ & $0.09(4)$  & $0.97(5)$  & $0.59$ &
$76.5\%$ \cr
$24$ & $6.33(25)$ & $1.43(8)$  & $0.62(3)$ & $0.12(4)$  & $0.99(6)$  & $0.86$ &
$55.0\%$ \cr
$26$ & $6.99(22)$ & $1.08(5)$  & $0.62(2)$ & $0.12(3)$  & $0.82(4)$  & $0.60$ &
$79.8\%$ \cr
$28$ & $7.21(28)$ & $1.05(6)$  & $0.60(3)$ & $0.14(4)$  & $0.82(5)$  & $0.72$ &
$70.6\%$ \cr
$30$ & $7.08(45)$ & $0.93(8)$  & $0.64(4)$ & $0.06(6)$  & $0.78(8)$  & $1.54$ &
$11.0\%$ \cr
\ \cr
\noalign{\hrule}}}$$
\vskip0.3 true cm
\centerline{\tabi}
\vskip13.0 true cm
\endinsert

\topinsert
$$\vbox{\halign{
\hfil#\hfil& \quad\hfil#\hfil&
 \quad\hfil#\hfil&
 \quad\hfil#\hfil& \quad\hfil#\hfil& \quad\hfil#\hfil&
 \quad\hfil#\hfil& \quad\hfil#\hfil\cr
\noalign{\hrule}
\ \cr
$\vk^2_{\rm min}$ & $\#$ data points &
 $z$ & $\alpha$ & $x$ & $y^2$ & $\chi^2/$d.o.f.\ & $Q$  \cr
\ \cr
\noalign{\hrule}
\ \cr
      &      & $17.2(249)$ & $0.48(5)$ & $-0.2(1)$  & $0.45(1)$ &        &
\cr
$0.5$ & $23$ & $12.8(20)$  & $0.45(7)$ & $0.11(17)$ & $0.45(6)$ & $0.98$ &
$48.1\%$ \cr
$0.3$ & $25$ & $13.4(11)$  & $0.43(4)$ & $0.17(8)$  & $0.42(3)$ & $0.94$ &
$53.8\%$ \cr
$0.5$ & $23$ & $11.5(2)$   & $0.49(1)$ & $0$        & $0.47(4)$ & $0.96$ &
$50.9\%$ \cr
$0.3$ & $25$ & $11.1(1)$   & $0.51(1)$ & $0$        & $0.39(2)$ & $1.22$ &
$21.7\%$ \cr
\ \cr
\noalign{\hrule}}}$$
\vskip0.3 true cm
\centerline{\tabinfi}
\vskip4.0 true cm

$$\vbox{\halign{
\hfil#\hfil& \quad\hfil#\hfil&
 \quad\hfil#\hfil&
 \quad\hfil#\hfil& \quad\hfil#\hfil& \quad\hfil#\hfil&
 \quad\hfil#\hfil& \quad\hfil#\hfil\cr
\noalign{\hrule}
\ \cr
$\vk^2_{\rm min}$ & $\#$ data points &
 $\Lambda_{\rm coul}^2$ & $z$ & $b' = {11 \over 12} b$ & $\alpha$ &
$\chi^2/$d.o.f.\ & $Q$  \cr
\ \cr
\noalign{\hrule}
\ \cr
$0.3$  & $25$ & $0.84(8)$  & $1.14(5)$ & $0.165(9)$ & $2.2(1)$  & $0.59$
 & $92.8\%$ \cr
$0.5$  & $23$ & $1.0(2)$   & $1.2(1)$  & $0.18(2)$  & $1.9(3)$  & $0.44$
 & $98.3\%$ \cr
$1.0$  & $20$ & $1.2(9)$   & $1.3(4)$  & $0.19(7)$  & $1.7(7)$  & $0.48$
 & $95.8\%$ \cr
\ \cr
\noalign{\hrule}}}$$
\vskip0.3 true cm
\centerline{\tabig}
\vskip5.0 true cm
\endinsert


\vfill\supereject\immediate\closeout\ffile{\parindent40pt
\baselineskip14pt\centerline{{\bf Figure Captions}}\nobreak\medskip
\escapechar=` \input figs.tmp\vfill\eject}

\topinsert
\vbox{
\vskip-3.0 true cm
\hskip-2.0 true cm
\centerline{
\epsfxsize=15.5 true cm
\epsfbox{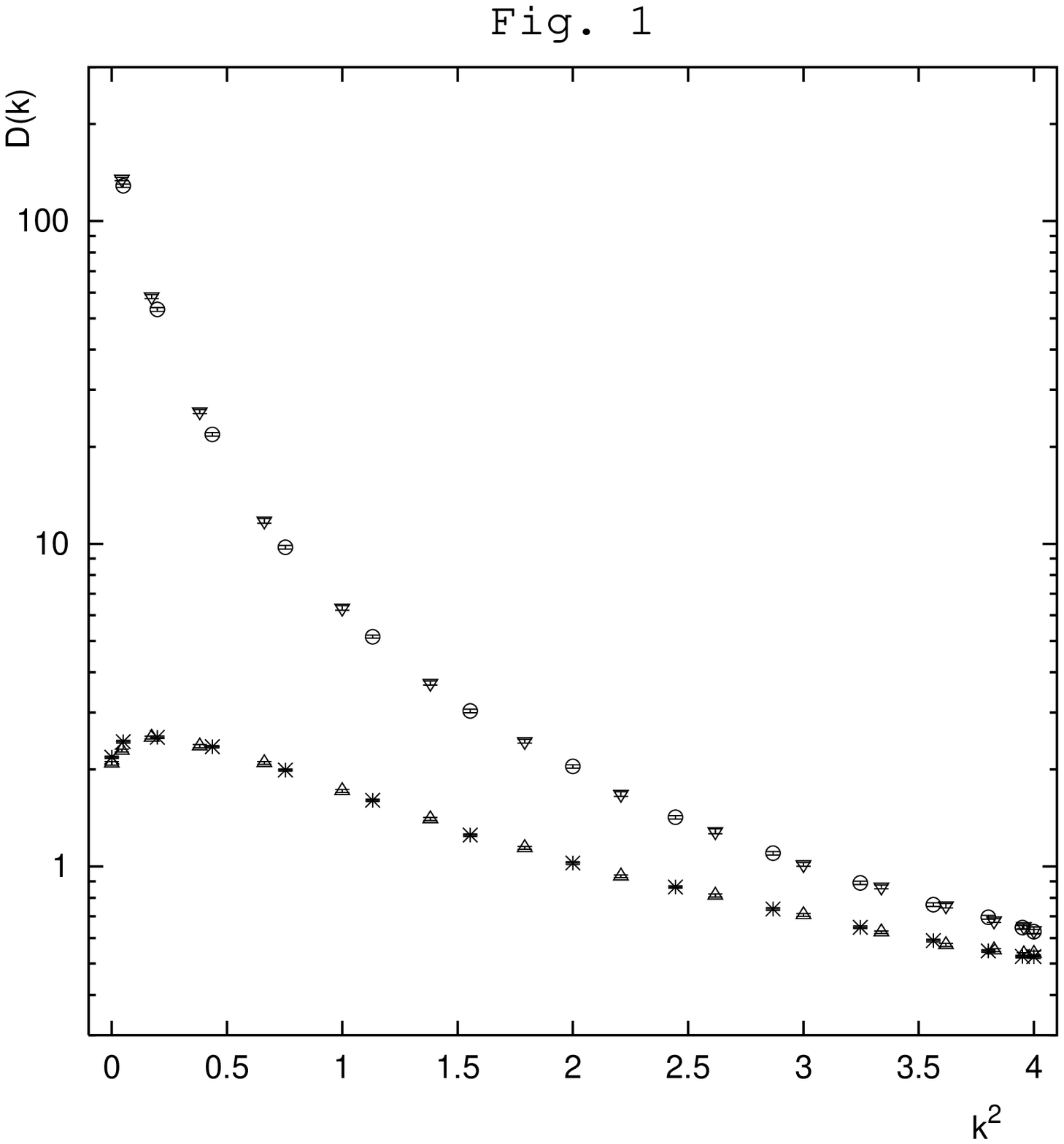}
}
\vskip2.0 true cm
}
\endinsert
\topinsert
\vbox{
\vskip-3.0 true cm
\hskip-2.0 true cm
\centerline{
\epsfxsize=15.5 true cm
\epsfbox{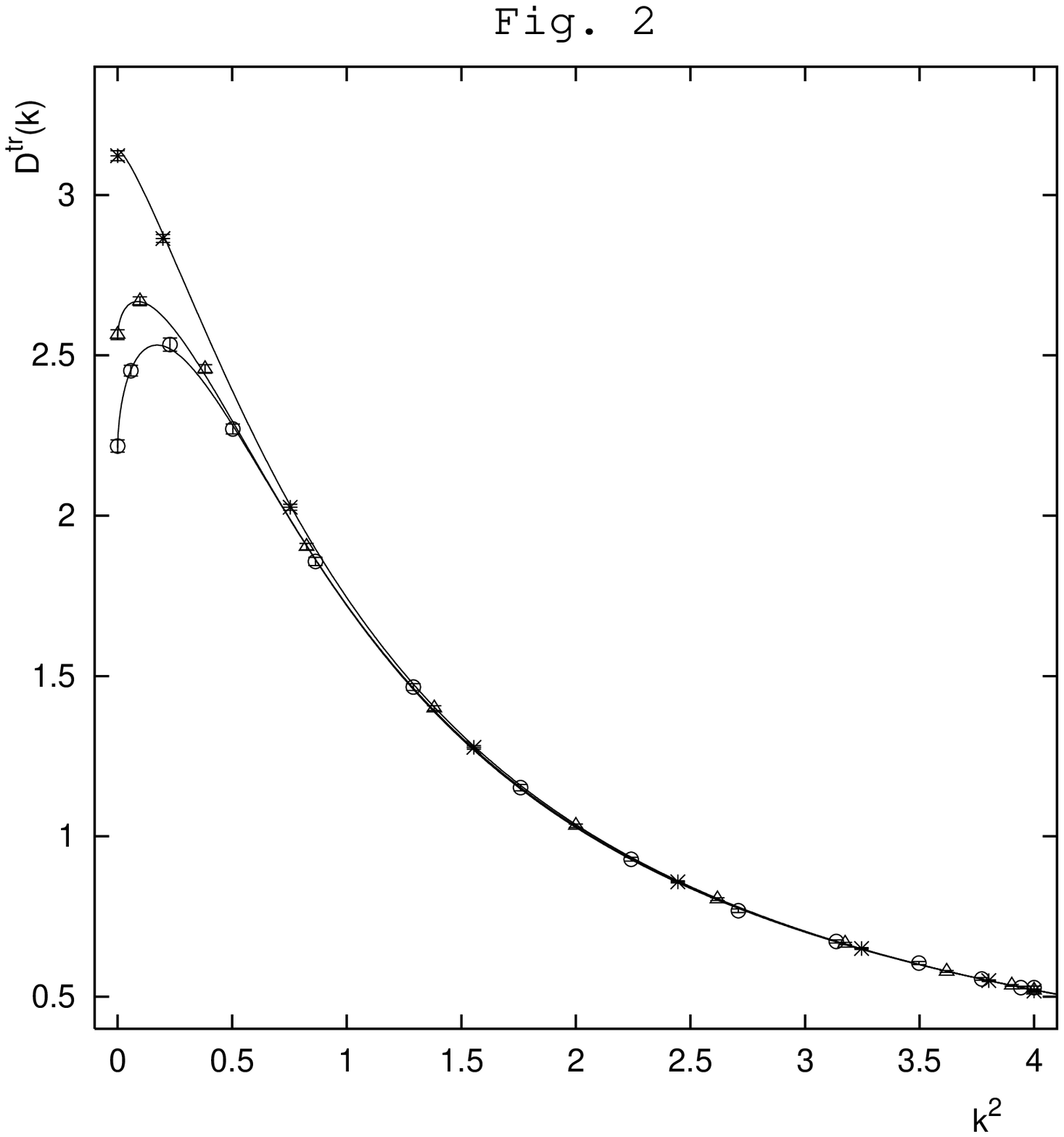}
}
\vskip2.0 true cm
}
\endinsert
\topinsert
\vbox{
\vskip-3.0 true cm
\hskip-2.0 true cm
\centerline{
\epsfxsize=15.5 true cm
\epsfbox{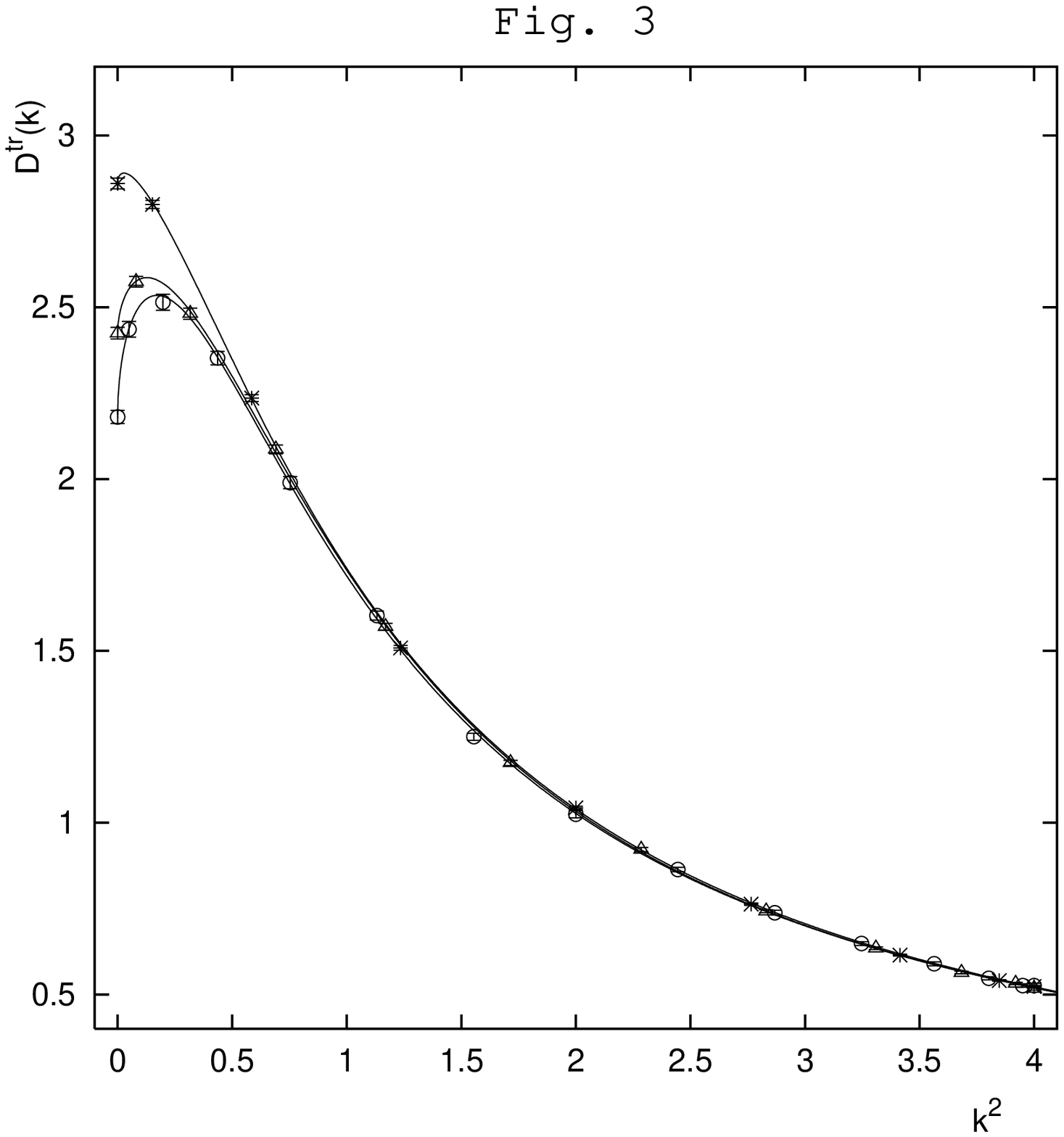}
}
\vskip2.0 true cm
}
\endinsert
\topinsert
\vbox{
\vskip-3.0 true cm
\hskip-2.0 true cm
\centerline{
\epsfxsize=15.5 true cm
\epsfbox{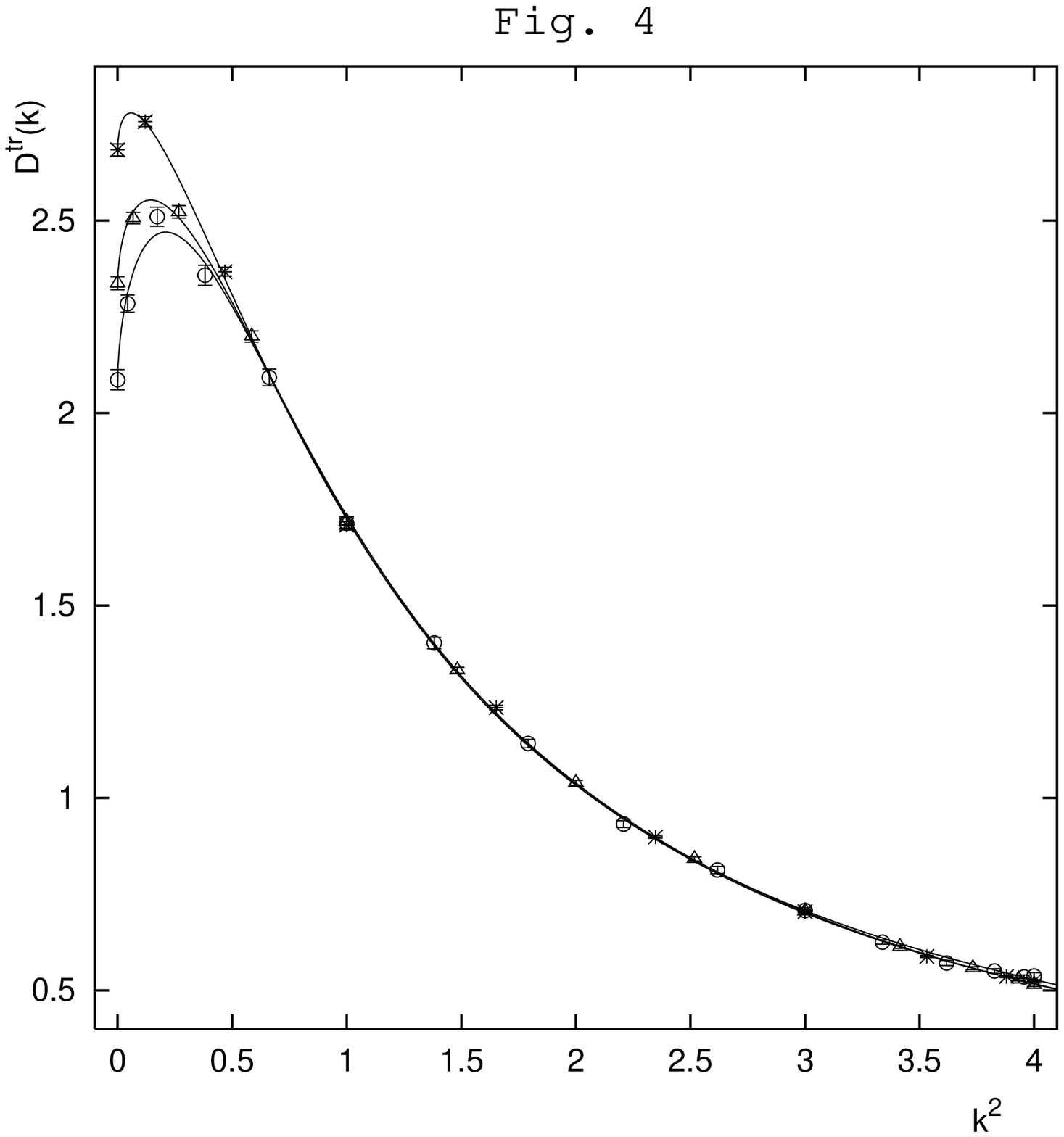}
}
\vskip2.0 true cm
}
\endinsert
\topinsert
\vbox{
\vskip-3.0 true cm
\hskip-2.0 true cm
\centerline{
\epsfxsize=15.5 true cm
\epsfbox{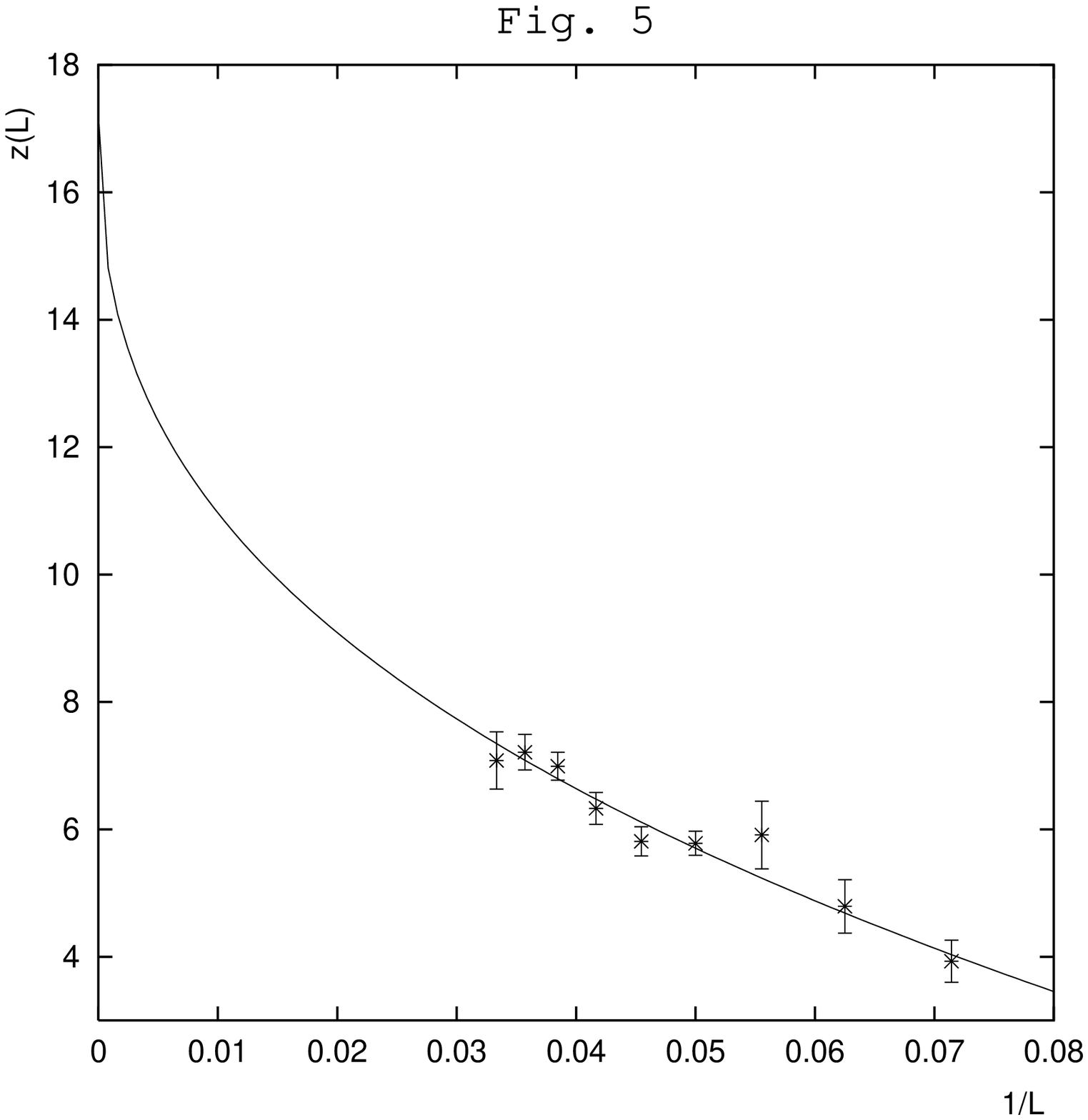}
}
\vskip2.0 true cm
}
\endinsert
\topinsert
\vbox{
\vskip-3.0 true cm
\hskip-2.0 true cm
\centerline{
\epsfxsize=15.5 true cm
\epsfbox{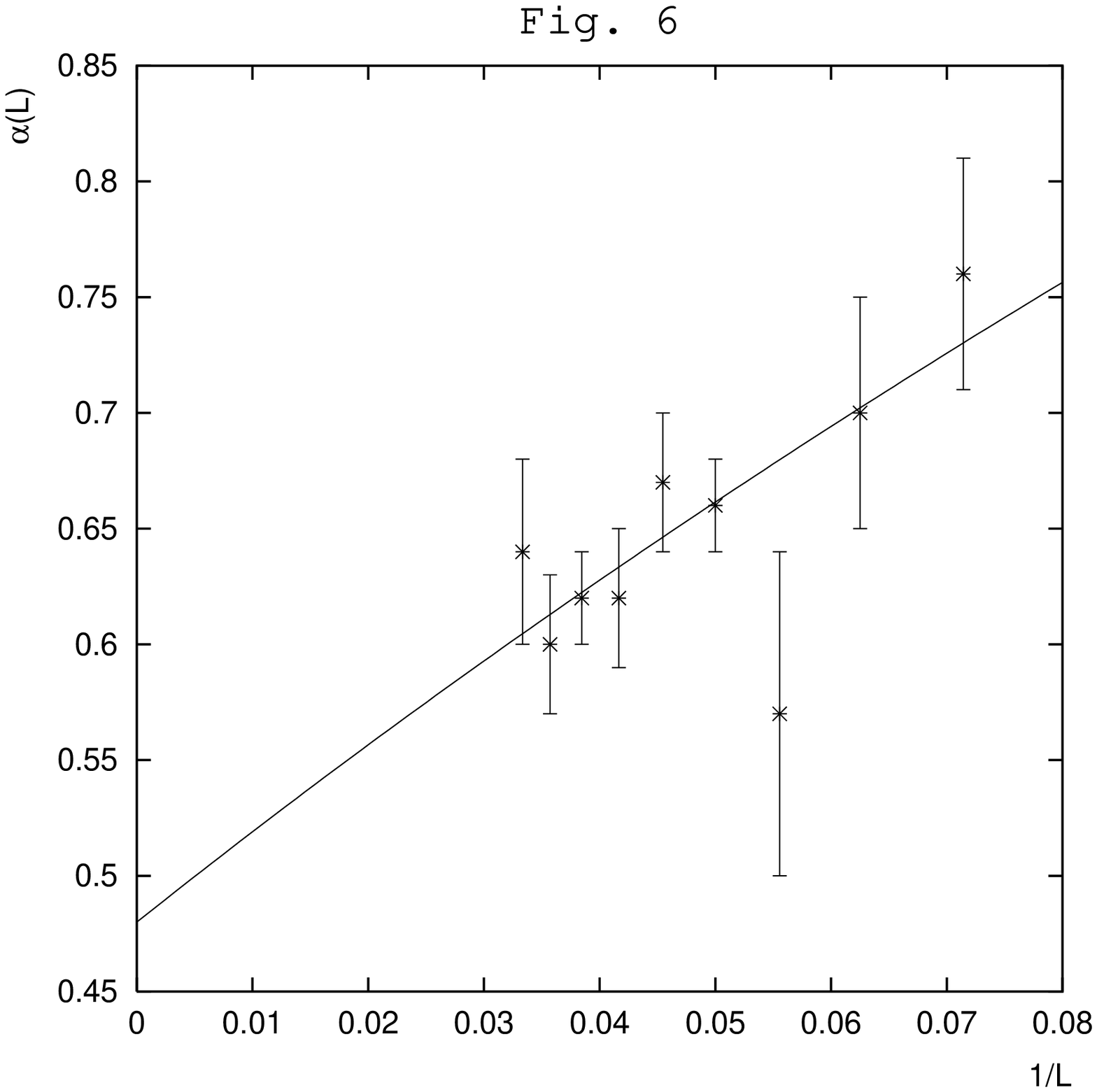}
}
\vskip2.0 true cm
}
\endinsert
\topinsert
\vbox{
\vskip-3.0 true cm
\hskip-2.0 true cm
\centerline{
\epsfxsize=15.5 true cm
\epsfbox{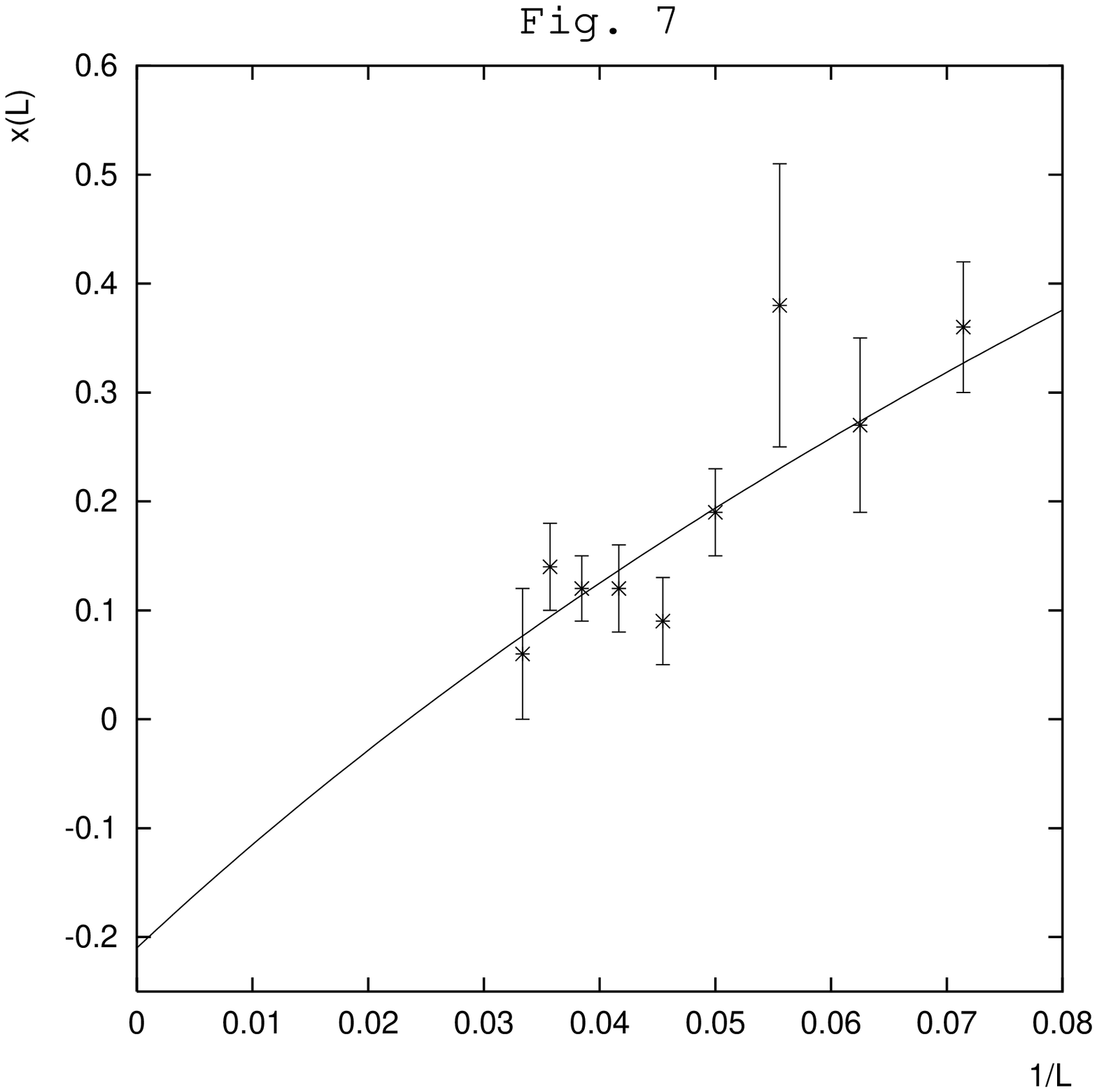}
}
\vskip2.0 true cm
}
\endinsert
\topinsert
\vbox{
\vskip-3.0 true cm
\hskip-2.0 true cm
\centerline{
\epsfxsize=15.5 true cm
\epsfbox{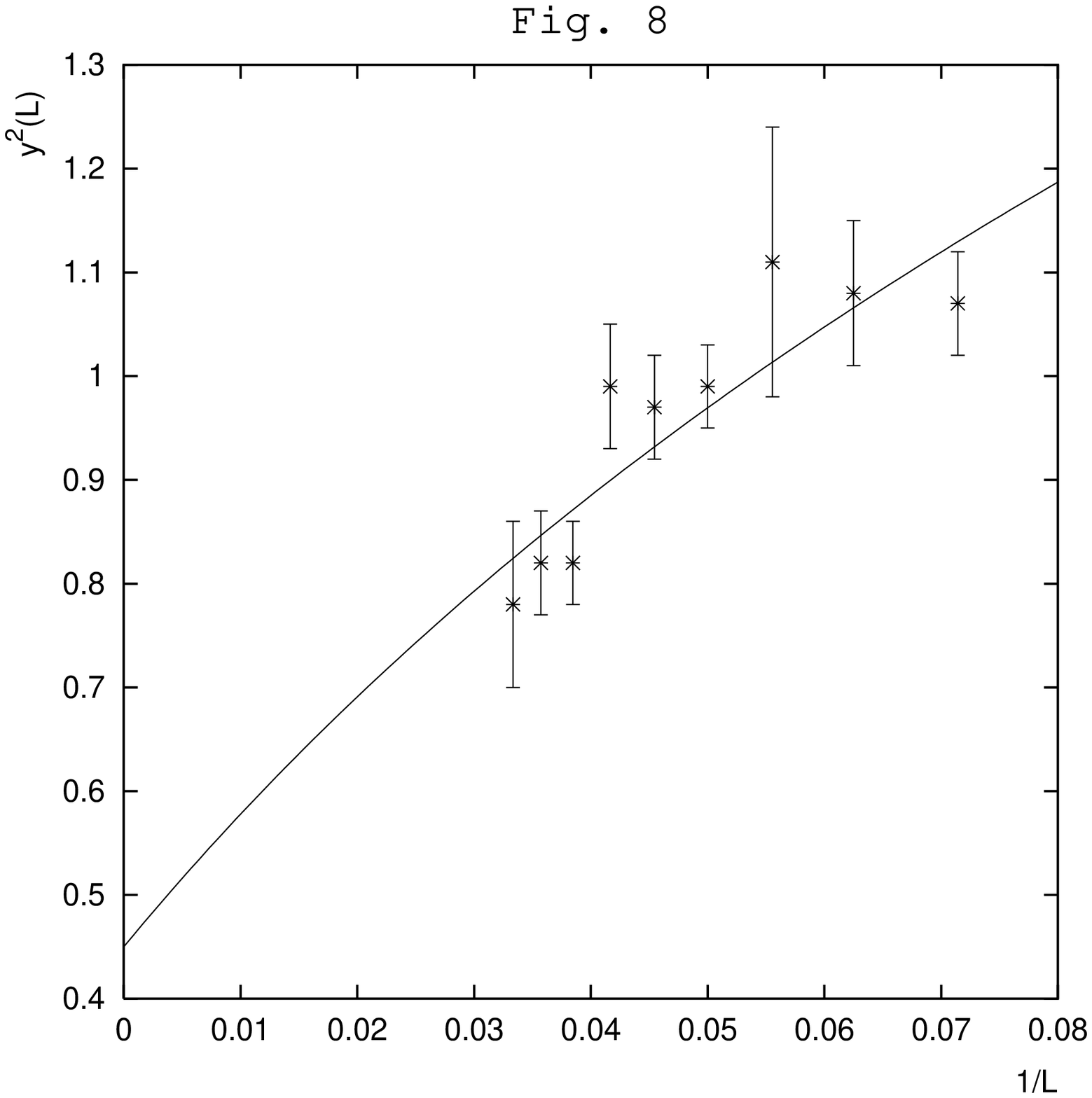}
}
\vskip2.0 true cm
}
\endinsert
\topinsert
\vbox{
\vskip-3.0 true cm
\hskip-2.0 true cm
\centerline{
\epsfxsize=15.5 true cm
\epsfbox{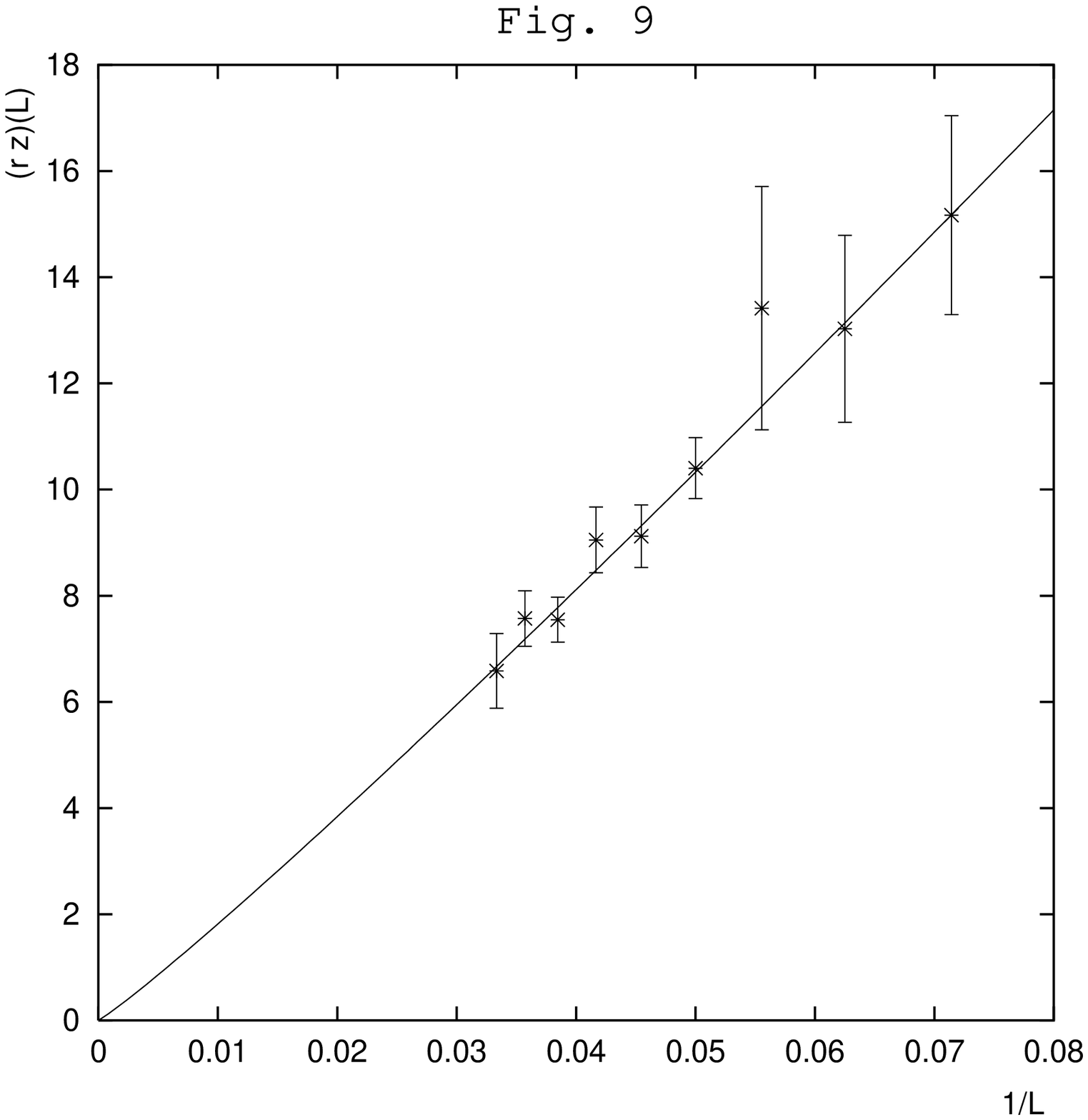}
}
\vskip2.0 true cm
}
\endinsert
\topinsert
\vbox{
\vskip-3.0 true cm
\hskip-2.0 true cm
\centerline{
\epsfxsize=15.5 true cm
\epsfbox{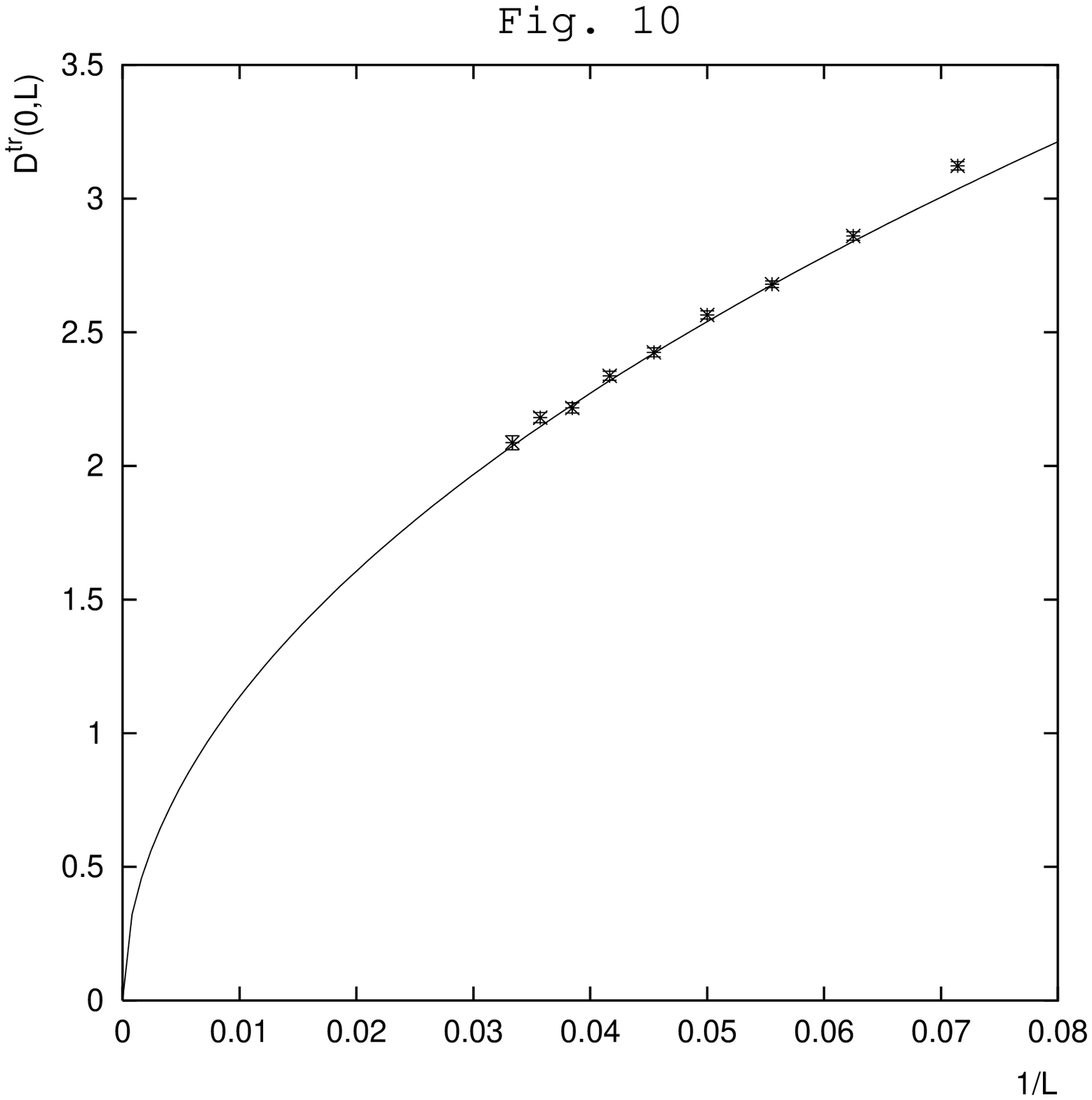}
}
\vskip2.0 true cm
}
\endinsert
\topinsert
\vbox{
\vskip-3.0 true cm
\hskip-2.0 true cm
\centerline{
\epsfxsize=15.5 true cm
\epsfbox{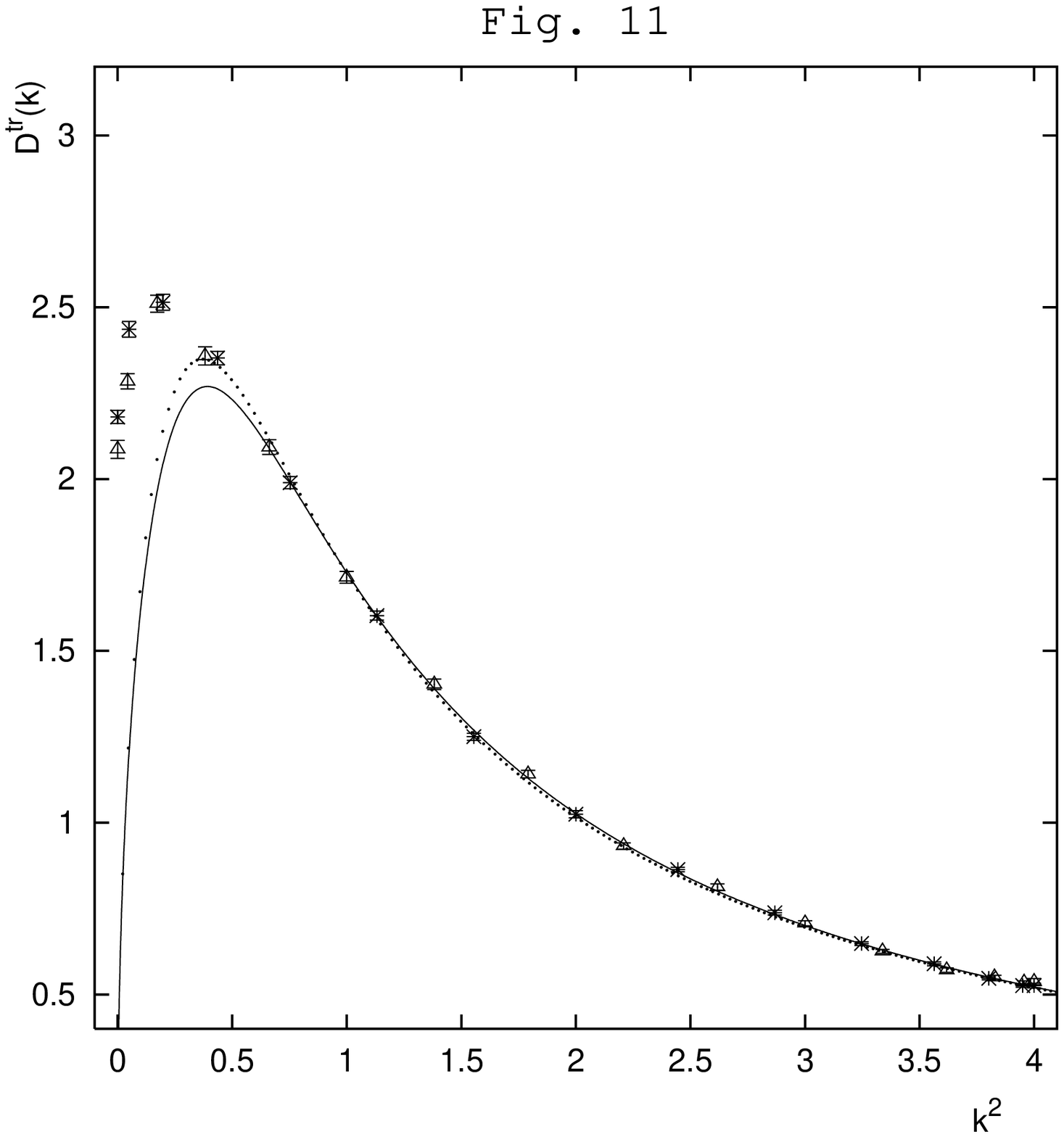}
}
\vskip2.0 true cm
}
\endinsert
\topinsert
\vbox{
\vskip-3.0 true cm
\hskip-2.0 true cm
\centerline{
\epsfxsize=15.5 true cm
\epsfbox{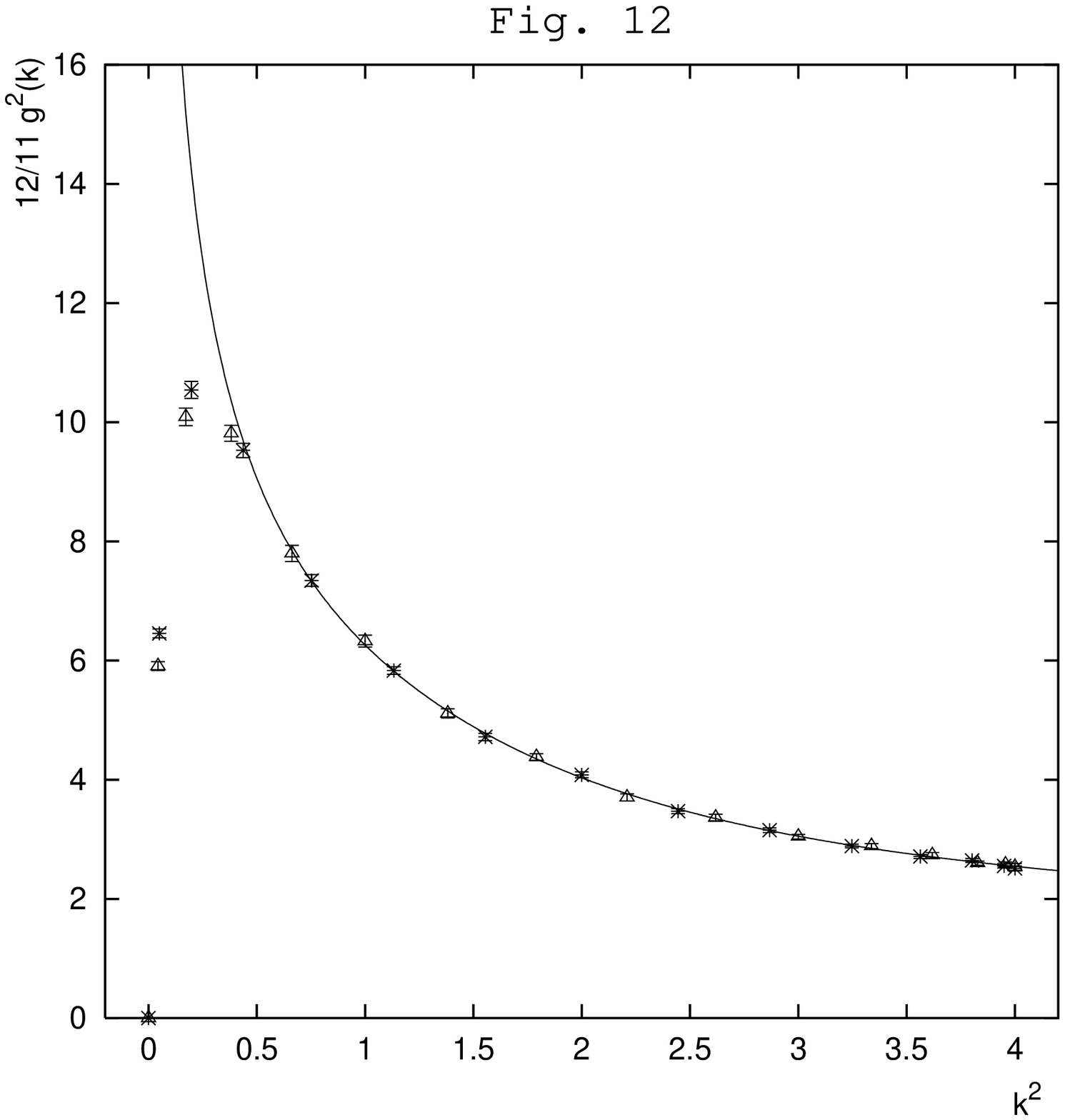}
}
\vskip2.0 true cm
}
\endinsert
%


\bye